# Complex-Orthogonal Gauge Manifolds and Simulated Scaling Laws for Single-Ended Calibration of Deforming Multimode Fibres

**Terry Wright[1,*], Callum J Wilson[1], and George S. D. Gordon[1]**

[1]Department of Chemical Engineering and Biotechnology, University of Cambridge, United Kingdom

[*]Corresponding author: tw567@cam.ac.uk

## Abstract

Imaging through a hair-thin multimode fibre requires its complex transmission matrix $U$, which is scrambled whenever the fibre bends, twists or drifts thermally. A calibrated reflector R using Lorentz reciprocity produces the round-trip matrix $Q = U^T RU$. We show that this is an exact Autonne–Takagi gauge quotient problem, in which $Q$ fixes $U$ up to a continuous complex-orthogonal gauge $O \in O(n, \mathbb{C})$ and a discrete signed-permutation branch group of order $n!\ 2^n$, with $U$ following in closed form once $O$ is known. When the transmission matrix is known, as at the calibration pins (fibre configurations for which U and Q are both known), whitened Hungarian column assignment recovers the branch exactly (relative Frobenius error F = $1.3 \times 10^{-15}$ over 39 simulated uniform bends of a 1 m fibre that differ only in radius), and the bends can be sorted from their round-trip measurements and transformed onto a single gauge manifold to F = $1.6 \times 10^{-15}$. There $O$ varies smoothly in the Lie algebra $\mathfrak{so}(n, \mathbb{C})$, so the gauge of a new configuration can be interpolated from neighbouring pins. For a live measurement, the branch and the bend position are recovered from the round trip alone, through the de-embedded reflector and the phases of the Takagi vectors, and the transmission matrix is reconstructed to a median F = $2.81 \times 10^{-4}$ with each bend withheld in turn and $3.95 \times 10^{-4}$ with half the bends withheld ($1.74 \times 10^{-4}$ and $2.66 \times 10^{-4}$ if the branch were taken from the true matrix). With only nine pins, reusing the best-matching stored matrix gives a median $F = 0.150$ (90th percentile 0.85), whereas keeping the live measurement's own Takagi factor, recovering its branch and interpolating only the gauge gives $5.5 \times 10^{-4}$ (0.47). Products of simulated pyMMF bend matrices show that the library needed for look-up grows by about 2.6× per independent bend. Two fibre models agree that look-up therefore fails beyond some length of free fibre but place it differently, at about 0.5 m in a phenomenological perturbative coupled-mode model and between 2 and 5 m in a mode-solver model of a step-index fibre, so its position depends on the fibre's bend sensitivity and has yet to be measured. In both models, interpolating only the gauge reconstructs the transmission matrix one to two orders of magnitude more accurately than look-up up to 1 m, and in the perturbative model mechanical constraint restores tractable library sizes. A spectral core-profile solver shows that a parabolic graded-index core slows the growth of the library with the number of compounded bends (5.7× fewer pins than a step-index core at eight bends when both are drawn from a common bend sampler). All errors are quoted with the optically irrelevant global phase removed, and the conclusions are tested against propagation-constant uncertainty. On measured transmission matrices of a 5 m step-index fibre, whose nearest recorded neighbours differ by $F \geq 0.67$ (median 0.94) and so are too far apart to test gauge interpolation, regularised reflector de-embedding of simulated round trips identifies the nearest stored calibration for 84% of held-out bends (93/111 with the regulariser fixed on separate development data).

## 1. Introduction

Multimode optical fibres (MMFs) transmit hundreds to thousands of optical modes through a cross-section narrower than a human hair, making them uniquely attractive for minimally invasive deep-brain neuroimaging, clinical micro-endoscopy, optical trapping and high-dimensional photonic processing[1−7]. When the forward transmission matrix (TM) $U \in \mathbb{C}^{n \times n}$ is known, wavefront shaping at the proximal facet can synthesise diffraction-limited focal spots or wide-field images at the distal tip[6−9]. The central obstacle to flexible MMF endoscopy is calibration stability: when the fibre bends, twists, heats or experiences mode-dependent loss, intermodal interference scrambles $U$, rendering a pre-insertion calibration useless precisely when distal optical access is impossible[8−15].

Reflection-mode calibration seeks to replace distal camera access with a known reflective element, beacon or guide star at or near the distal facet[10,16−21]. Gu, Mahalati and Kahn proposed a flexible MMF endoscope with a partial distal reflector enabling proximal-only recalibration after bending[16], and subsequently used reflected-to-transmitted perturbation correlations to improve distal spot formation[22]. Because Lorentz reciprocity enforces transposition on the backward propagation path, the proximal round-trip reflection matrix takes the complex-symmetric form

$$Q = U^T R U, \qquad Q^T = Q \in \mathbb{C}^{n \times n},$$

where $R = R^T \in \mathbb{C}^{n \times n}$ is the modal reflection matrix of the distal end. In Gu et al.'s formulation[16], recovery of $U$ is made tractable by assuming that the fibre transmission matrix is strictly unitary ($U^H U = I_n$) and lossless. Under unitarity, the round-trip ambiguity collapses to a discrete diagonal sign choice that can be tracked from a close prior estimate of $U$.

In realistic endoscopic fibres, however, $U$ is significantly non-unitary due to bend-induced radiation into cladding modes, numerical-aperture truncation, mode-dependent attenuation and single-polarisation detection[9,10,19]. Gordon et al. removed the unitary assumption by introducing a distal stack of wavelength-dependent metasurface reflectors[10]: when three reflector matrices $R(\lambda_i)$ possess distinct eigenvalue spectra and the fibre TM varies slowly across the characterisation wavelengths, the simultaneous equations $Q(\lambda_i) = U^T R(\lambda_i) U$ algebraically eliminate the non-unitary ambiguity, with a first-order iterative solver compensating for chromatic TM drift[10]. Zheng et al subsequently recast multi-wavelength reflection recovery as a neural inverse problem using a phase-insensitive loss to handle residual global phase gauge[19], while Wen et al. tracked bend-induced changes at up to 1 kHz using the distal-facet Fresnel reflection as a spatial-frequency beacon[20]. In parallel, recent database and deep-learning approaches have attempted to bypass distal access by recording large look-up libraries of bent fibre states and retrieving or interpolating stored TMs from proximal measurements[13,14,21].

In this work rather than assuming unitarity or relying on brute-force look-up tables we determine the exact mathematical structure of the information left unresolved by a single-ended reflection measurement $Q = U^T RU$, and also how these unresolved degrees of freedom evolve as the fibre is bent.

The Autonne–Takagi factorisation[23−25]provides a way to work with the complex-symmetric matrices Q and R. Every complex-symmetric matrix $A = A^T \in \mathbb{C}^{n\times n}$ admits a unitary congruence $A = U_A \Sigma_A U_A^T$ with unitary $U_A \in U(n)$ and non-negative diagonal $\Sigma_A$ containing the singular values of $A$. Applying Takagi factorisation to $Q = U^T RU$ fixes the right Takagi square-root factor $B_Q = D_Q^{1/2} V^T$ directly from the live proximal measurement. However, it leaves a continuous complex-orthogonal gauge $O \in O(n, \mathbb{C}) = \{O \in \mathbb{C}^{n\times n}: O^T O = I_n\}$ unresolved, modulo a discrete $n!\,2^n$ signed-permutation branch ambiguity $G = PS$, (where P permutes the columns and S changes their signs). Because the factorisation is unique only up to a signed permutation, independent numerical Takagi solvers can land on any one of the $n!\,2^n$ signed-permutation sheets at nearby bends[25,26]; where singular values coincide the variation in the columns of V can also lose smoothness[25]. Our approach is to remove discrete signed-permutation jumps G = PS by whitened column-correction matching across sorted calibration pins. (sample bends where U, Q, R are known) If we imagine a set of bends determined by a set of parameters $\boldsymbol{\theta}$, then the gauge $O(\boldsymbol{\theta})$ traces a smooth, low-dimensional **aligned gauge manifold** inside the complex skew-symmetric Lie algebra $\mathfrak{so}(n, \mathbb{C})$. The gauge $O(\boldsymbol{\theta})$ contains the extra information above Q that is required to estimate $\widehat{U}$. Each pin is a sample point on the aligned gauge manifold, and with enough samples its structure can be learned, allowing for interpolation via Lie algebra tangent coordinates. Interpolating the missing gauge $\widehat{O(\boldsymbol{\theta})} \in O(n, \mathbb{C})$ in $\mathfrak{so}(n, \mathbb{C})$ while keeping the newly measured round-trip factor $B_{Q,new} = D_{Q,new}^{1/2} V_{new}^T$ enables the reconstruction of transmission matrix $\widehat{U}$, achieving orders-of-magnitude lower error than averaging raw transmission matrices in Euclidean space. Treating a deforming MMF of length $L$ as a multiplicative cascade of coupled-mode transfer matrices gives a scaling law, which we test in two fibre models, for the required number of calibration pins $N_{pin}(L)$, explaining why naive global look-up becomes exponentially expensive in the number of independent bends, and showing how mechanical joint constraints, full $2N \times 2N$ vectorial polarisation control, structured distal reflectors with multi-wavelength dispersion achieve high quality focus via single-ended tracking. Once the aligned gauge manifold has been learned for a fibre, its transmission matrix $U$ can be estimated from a live, new round-trip measurement $Q_{live}$. Each stored pin matrix $U_j$ ,represented by its corresponding pair $(O_j,\ B_j)$, can be 'de-embedded' from $Q_{live}$ to estimate the reflector $\hat{R} = U_j^{-T} Q_{live} U_j^{-1}$ . Comparing these estimates with the calibrated reflector $R$ identifies candidate neighbouring pins. Pins with smaller reflector mismatch receive greater weight when interpolating their local logarithmic gauge coordinates. The resulting gauge estimate $\hat{O}$ is then combined with the live Takagi factor $B_{live}$ to reconstruct $\widehat{U}$ . This supports cold-start estimation without a previous-state estimate, provided the scores identify the appropriate manifold region and the calibration pins adequately sample it. One limitation of using $\hat{R}$ as a score to find $\hat{O}$ is that it cannot distinguish cases where Q is constant and only O changes; however, this does not appear to limit its usefulness in gauge interpolation. With a lossless fibre there is no gauge to interpolate – the gauge is a fixed sign and reordering of the modes. Gauge interpolation is necessary because real fibres are not unitary.

## 2. Results

Three quantities are used throughout. The relative Frobenius error minimised over a global phase, $F = \min_{\varphi} \parallel e^{i\varphi}\hat{U} - U \parallel_F / \parallel U \parallel_F$, measures reconstruction accuracy directly. The global phase is removed because it cannot change any focus, while every measured matrix and every simulated propagation carries its own (Appendix H.1); $F = \sqrt{2}$ means no information about $U$ survives. The predicted focus quality $\eta = max(0,1 - F^2/2)^2$ rescales $F$ onto [0,1] and equals the squared normalised overlap $\left|tr(\hat{U}^{\dagger}U)/n\right|^2$ when $\parallel \hat{U} \parallel_F = \parallel U \parallel_F$.

It allows us to relate errors in the reconstruction of U to the optical focus that can be achieved through the fibre. The phase-conjugation Strehl ratio $S_p$ is the directly evaluated focusing quantity: the intensity delivered to a chosen target point when the input is calculated from the estimated transmission matrix $\hat{U}$, divided by its ideal value when $\hat{U} = U$, for equal input power. $S_p$ is analogous to the Strehl ratio in classical optics, which compares the peak intensity of an aberrated point-spread function with its diffraction-limited value.

η is a coherent measure over the whole matrix—the squared magnitude of the normalised complex overlap—whereas $S_p$, averaged over targets, is a mean of intensity ratios. Supplementary Fig. S1 shows that they agree approximately under the tested error conditions, so η can be used as a proxy for focus quality, although the two are not generally equal. A Strehl ratio $S_p \geq 0.80$ is conventionally taken as the diffraction-limited criterion. We retain the dotted reference line at η = 0.80 on the plots as a useful indication of likely good focusing, while recognising that it does not establish the exact diffraction-limited threshold for every target.

### 2.1 Autonne–Takagi quotient geometry, exact branch recovery and cross-bend program sorting

Let $U \in GL(n, \mathbb{C})$ denote the forward transmission matrix from proximal to distal fibre modes and $R = R^T \in GL(n, \mathbb{C})$ the calibrated distal reflector. Since $R$ and the measured proximal round-trip matrix $Q = U^T R U$ are complex symmetric, both admit Autonne–Takagi factorisations[23−25]:

$$R = W D_R W^T, \qquad Q = V D_Q V^T,$$

where $W, V \in U(n)$ are unitary and $D_R = diag\big(d_{R,1}, \dots, d_{R,n}\big)$, $D_Q = diag\big(d_{Q,1}, \dots, d_{Q,n}\big)$ are real non-negative diagonal matrices. Defining the square-root factors

$$A = W D_R^{1/2}, \qquad B = V D_Q^{1/2},$$

so that $R = AA^T$ and $Q = BB^T$ (equivalently $A_R = A^T$ and $B_Q = B^T$, so that $R = A_R^T A_R$, $Q = B_Q^T B_Q$ and $O = A_R U B_Q^{-1}$; this form is used below), any transmission matrix $U$ satisfying $Q = U^T R U$ defines a matrix

$$O = A^T U B^{-T} = D_R^{1/2} W^T U \big(D_Q^{1/2} V^T\big)^{-1} \in O(n, \mathbb{C}),$$

which satisfies complex orthogonality identically:

$$O^T O = B^{-1} U^T A A^T U B^{-T} = B^{-1} Q B^{-T} = I_n.$$

Solving for $U$ yields the exact single-ended reconstruction law (Fig. 1a):

$$\widehat{U}(O) = A^{-T} O B^T = (W^T)^{-1} D_R^{-1/2} O D_Q^{1/2} V^T = \overline{W} D_R^{-1/2} O D_Q^{1/2} V^T.$$

However, the Autonne–Takagi factorisation of $Q$ is not unique. For any signed-permutation matrix

$$G = PS \in O(n, \mathbb{R}), \qquad G^T G = I_n,$$

where $P$ is a $n \times n$ permutation matrix and $S = diag(\pm 1, \dots, \pm 1)$, the transformed Takagi factors $V_1 = VG$ and $D_{Q,1} = G^T D_Q G$ leave $Q = V_1 D_{Q,1} V_1^T$ unchanged while transforming the inferred gauge via $O_1 = OG$. Consequently, the discrete branch group $G$ ($|G| = n!\,2^n \approx 3.4 \times 10^{79}$ for $n = 50$) replicates a smooth physical deformation manifold into $n!\,2^n$ disjoint parallel sheets (Fig. 1b).

To recover the signed permutation and place every Takagi output on one **aligned gauge manifold**, we work in whitened coordinates. Since $U = A^{-T} O B^T$, the matrix $U B^{-T} = A^{-T} O$ depends on the Takagi branch only through the columns of $O$. For an already-aligned reference pin, $X = A^{-T} O_{\mathrm{ref}}$; for a new pin whose factorisation lies on branch $G$, $Y = A^{-T} O_{\mathrm{new}} = A^{-T} O_{\mathrm{ref}} G$ (exactly for the same fibre state, approximately for a neighbouring one), so $X \approx Y G^T$. We normalise the columns of $X$ and $Y$ and solve the $O(n^3)$ Hungarian linear assignment problem on the absolute cross-correlation matrix $\big|C_{kj}\big| = \big|\hat{X}_{:,k}^H \hat{Y}_{:,j}\big|$, with signs $\hat{s}_k = sign\big(Re\, C_{k,\hat{\pi}(k)}\big)$. This returns $\hat{G} \approx G$, the branch itself, and the correction $O_{\mathrm{new}} \hat{G}^T = O_{\mathrm{ref}}$ places the new pin on the reference sheet.

We conducted 2 progressively more demanding tests across the 40 saved pyMMF[28] bend transmission matrices (uniform bends in a single plane that differ only in radius; $n = 50$ modes, $L = 1.0$ m, bend radii $R_{bend} \in [0.20, 4.00]$ m, Fig. 1c).

Bend 1 ($R_{bend} = 0.20$ m) was excluded from the test as it is an explicit extreme-loss stress case: severe bend radiation gives $cond(U) = 1.567 \times 10^6$, $cond(Q) = 2.032 \times 10^{12}$ and a minimum Takagi value of $9.39 \times 10^{-9}$, at the double-precision conditioning limit. Across the remaining 39 well-conditioned bends (`2.mat–40.mat`, $R_{bend} \in [0.30, 4.00]$ m, $maxcond(Q) = 2.70$) a random Takagi branch $G_b$ destroys the reconstruction completely, with median $F = 1.4434$ and $\eta = 0$, whereas whitened Hungarian branch recovery returns a median $F = 1.339 \times 10^{-15}$ (maximum $1.579 \times 10^{-15}$, minimum column-match score 1.0). Branch recovery is therefore exact, not approximate.

Genuine calibration, however, requires a mechanical program of bends in which $(U_k, Q_k)$ is recorded at each pin with an unknown branch $G_k$, the bends are sorted automatically, and each new pin is pulled onto the manifold using the nearest already-aligned gauge $O_{k-1}$ (Fig. 1c). Greedy nearest-neighbour sorting on the proximal round-trip distance $d_Q(i,j) = \| Q_i - Q_j \|_F / \| Q_i \|_F$ orders the 39-bend trajectory when the walk starts from the loose end of the programme ($r = 0.999$), and cross-bend Hungarian matching against $O_{k-1}$ places every pin on the same sheet: the maximum sheet discrepancy is $1.613 \times 10^{-15}$ and the pin reconstruction error $5.2 \times 10^{-16}$. The three estimates of $O_k$ in Fig. 1c are scored by the same relative Frobenius error, leaving the branch uncorrected gives $F_O = 1.414$ (red circles), which carries no information about $O_k$; taking the gauge of the nearest already-aligned pin gives a median $F_O = 1.10 \times 10^{-3}$ (orange triangles); and Lie-interpolating $\hat{O}_{Lie,k}$ in $\mathfrak{so}(n, \mathbb{C})$ from the neighbouring pins, leave-one-out, gives a median $F_O = 1.735 \times 10^{-4}$ (green diamonds), propagating to $F_U = 1.736 \times 10^{-4}$.

Mapping the sequentially aligned gauges to the complex skew-symmetric Lie algebra $\mathfrak{so}(n, \mathbb{C}) = \{L \in \mathbb{C}^{n \times n} : L^T = -L\}$ via $L_k = \Pi_{\mathfrak{o}}\big(\log(O_0^{-1} O_k)\big) = \frac{1}{2}(\log(O_0^{-1} O_k) - log(O_0^{-1} O_k)^T)$ confirms exact Lie-group closure: the median raw-logarithm skew-symmetry residual is $1.05 \times 10^{-13}$ and the re-exponentiation residual $\| O_0 \exp(L_k) - O_k \|_F / \| O_k \|_F$ is $1.67 \times 10^{-14}$. Holding out every second bend (19 train / 19 test bends, Fig. 1d), full-coordinate Lie PCHIP interpolation achieves median $F = 2.655 \times 10^{-4}$ ($\eta = 1.000$) and tangent-PCA PCHIP ($d = 8$ components) achieves $F = 2.767 \times 10^{-4}$, outperforming nearest-pin gauge reuse ($F = 9.941 \times 10^{-4}$) by $3.7 \times$ and LOO-CV RBF kernel ridge regression ($F = 6.65 \times 10^{-3}$) by $25 \times$. These held-out errors, here and in Fig. 1c,d, take the Takagi branch of each withheld bend from its own transmission matrix, which a live measurement does not supply (oracle alignment). Section 2.6 finds the branch and the bend position from Q alone; for the same split this gives a median F = $3.95 \times 10^{-4}$ ($2.81 \times 10^{-4}$ leave-one-out), and only the tightest withheld bend (R = 0.39 m) falls below η = 0.80. Adding random symmetric noise to every measured $Q$ raises the held-out median error smoothly, from $F = 2.2 \times 10^{-4}$ at $\epsilon_Q = 10^{-6}$ to $F = 0.046$ at $\epsilon_Q = 10^{-3}$ (η = 0.998). An error in the reflector has no effect if the same $R$ is used to build the pins and to reconstruct, since it cancels exactly; if the reflector changes after calibration, a relative change $\epsilon_R = 10^{-3}$ gives $F = 0.046$.

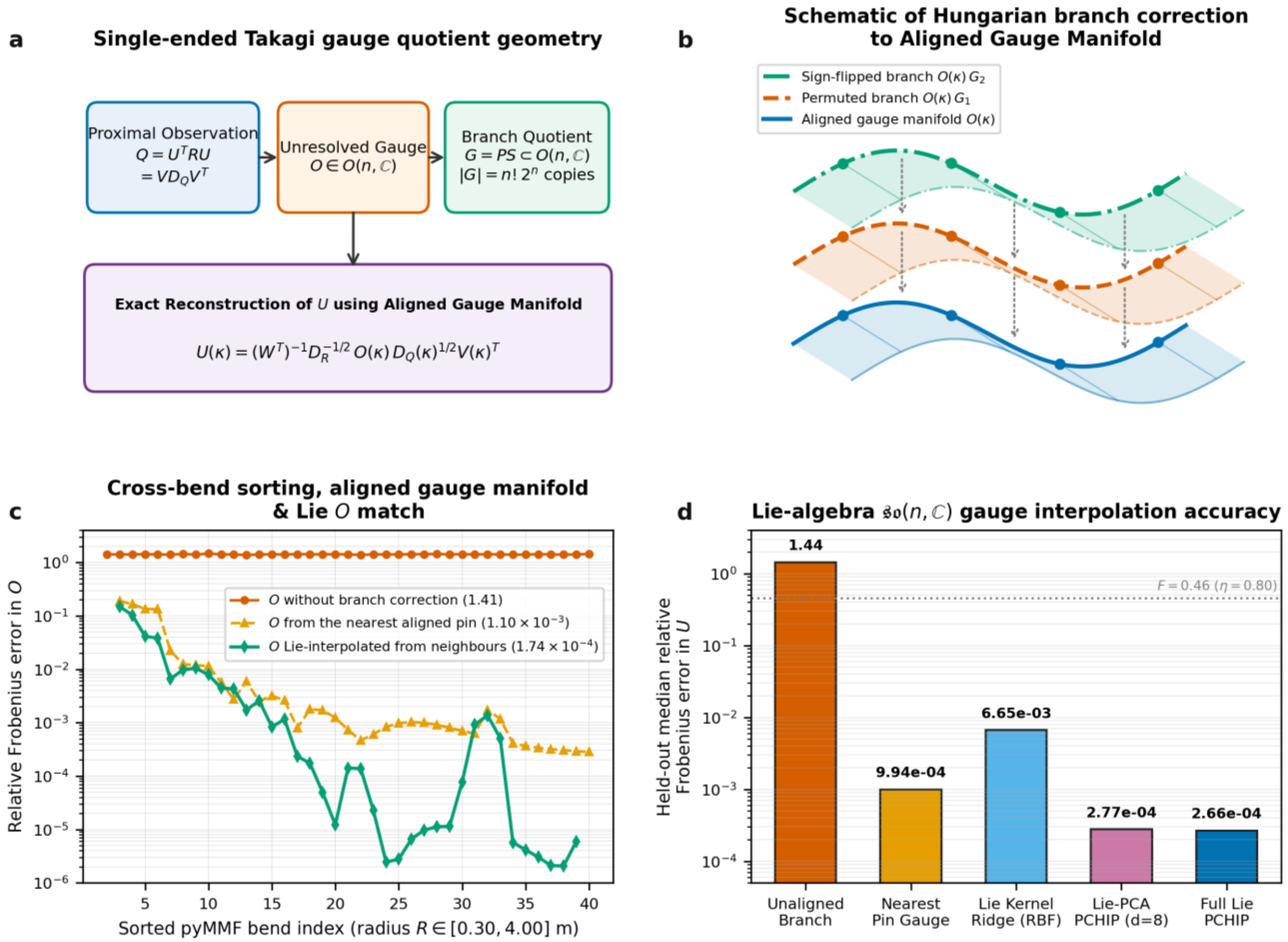

**Fig. 1.** Single-ended Autonne–Takagi gauge geometry and manifold alignment. **a**, A single-ended measurement fixes $Q = U^{\top}RU = VD_QV^{\top}$ but leaves a complex-orthogonal gauge $O \in O(n, \mathbb{C})$ undetermined, with the Takagi branch group $G = PS$ adding $n!\,2^n$ discrete copies; $U$ follows exactly once $O$ is fixed. **b**, Schematic: branches lie on parallel translated sheets, and whitened Hungarian column assignment against an already-aligned neighbouring pin selects the correct one. **c**, Applied to 39 pyMMF bends ($n = 50$, $R \in [0.30, 4.00]$ m), the relative Frobenius error in $O$ falls from 1.41 with no branch correction to $1.10 \times 10^{-3}$ using the nearest aligned pin, and to $1.74 \times 10^{-4}$ when $O$ is Lie-interpolated in $\mathfrak{so}(n, \mathbb{C})$ from neighbouring pins. Chaining the correction pin-to-pin across all 39 bends reproduces direct alignment to $1.61 \times 10^{-15}$. **d**, With every pin on one sheet, held-out $U$ is recovered to a median relative Frobenius error of $2.66 \times 10^{-4}$ by full Lie PCHIP interpolation, against 1.44 with no branch correction. The dotted line marks $F = 0.46$, where the predicted focus quality $\eta = (1 - F^2/2)^2$ equals 0.80 (Fig. S1). In c and d each held-out bend's Takagi branch is taken from its own transmission matrix (oracle alignment); finding it from Q alone (Section 2.6) gives medians of $2.81 \times 10^{-4}$ (leave-one-out) and $3.95 \times 10^{-4}$ (Fig. S3b).

## 2.2 Bend Hamiltonians and gauge composition

To explain physically why the aligned gauge $O(\kappa)$ is smooth on a single bend yet becomes progressively harder to sample across multiple bends, we derived the conformal-mapping coupled-mode Hamiltonian of a bent MMF segment of length $\Delta z$, curvature $\kappa = 1/R_{bend}$ and bend-plane azimuth $\phi$ (Appendix D.1):

$$H(\kappa,\phi) = diag(\beta_0 + i\gamma_0) + \kappa\big(\cos\phi\, C_x + sin\phi\, C_y\big) + \kappa^2\big(\cos^2\phi\, Q_{xx} + \sin^2\phi\, Q_{yy} + sin(2\phi)\, Q_{xy}\big) + \mathcal{O}(\kappa^3).$$

Here $C_x, C_y$ mediate first-order dipole coupling between adjacent principal mode groups ($|\Delta g| = 1, |\Delta \ell| = 1$) while $Q_{xx}, Q_{yy}, Q_{xy}$ capture second-order geometric and elasto-optic wavefront compression ($|\Delta g| \in \{0,2\}$). Fitting polynomial generator models of order $p \in \{1,2,3,4\}$ across odd pyMMF bends and predicting held-out even bends (Fig. 2a) improves the median held-out relative Frobenius error in $U$ from $F = 0.3247$ ($\eta = 0.8973$) at linear order ($\kappa H_1$) to $F = 0.2683$ ($\eta = 0.9293$) at quadratic order ($+\kappa^2 H_2$) $F = 0.2581$ ($\eta = 0.9345$) at third order and $F = 0.236$ ($\eta = 0.945$) at fourth order with the global phase removed. Each order helps, but the increments ($-0.056$, $-0.010$, $-0.022$) do not decay, so the fourth-order model is a usable fit rather than a convergent expansion. Because the bend Hamiltonian depends smoothly on curvature, a single bend traces a smooth one-dimensional curve $O(\kappa)$ in $SO(n,\mathbb{C})$, and hence a smooth curve $L(\kappa)$ in its Lie algebra $\mathfrak{so}(n,\mathbb{C})$, whose tangent line $O^{-1}dO/d\kappa$ turns slowly with $\kappa$ (participation ratio ≈ 1.3); this holds wherever the Takagi values of Q remain distinct and U is non-unitary, and it is what allows the gauge to be interpolated from a few pins.

In the perturbative model (Appendix A.4), when $M$ independent bent segments of length $\ell_c = 0.10$ m are concatenated in series ($U^{(M)} = \prod_{m=1}^{M} \exp\big(i\ell_c H(\kappa_m, \phi_m)\big)$, Fig. 2b,c), non-commuting Baker–Campbell–Hausdorff (BCH) cross-terms $[H_m, H_j] \neq 0$ cause the Lie-algebra velocity magnitude in $\mathfrak{so}(n,\mathbb{C})$, $v(M) = \| \big(O^{(M)}\big)^{-1} \partial_s O^{(M)} \|_F$, to grow by three orders of magnitude in the program coordinate $s$ from $v = 0.021$ at $M = 1$ ($L = 0.1$ m) to $v = 20.84$ at $M = 12$ ($L = 1.2$ m), generating localized high-curvature ("spiky") transitions where the 95th-percentile-to-median velocity ratio $P_{95}(v)/med(v)$ peaks at 8.47 ($M = 4$, Fig. 2c). The motion nevertheless stays low-dimensional (participation ratio at most 1.84; Fig. 2c). This is consistent with the 9-pin benchmark on the pyMMF bends (Fig. 2d): the median error is $F_{med} = 2.28 \times 10^{-3}$, but every held-out bend that fails $\eta \geq 0.80$ lies at $R \leq 0.98$ m, where the pins are 0.49 m apart, so the tail comes from leaving the rapid small-radius transition unpinned.

To evaluate practical single-ended reconstruction where the bend coordinate is unknown and candidate pins must be selected purely from the new proximal measurement $Q_{new}$, we compared six algorithms under identical de-embedded reflector scoring,

$$\hat{R} = U_j^{-T} Q_{new} U_j^{-1}$$

$$\epsilon_R(j) = \frac{\| \hat{R} - R \|_F}{\| R \|_F}$$

using 9 space-filling calibration pins across the 39 pyMMF bends (Fig. 2d). This score needs no ground truth: for each stored pin $U_j$ it de-embeds the distal reflector from the new measurement $Q_{new}$, and asks how closely the recovered reflector matches the $R$ that was actually fabricated and characterised. The six methods then split into two families by what they do with that score. **Retrieval methods** use it to choose: the best-scoring pin wins and its stored matrix is returned. **Interpolation methods** use the scores of the top $k$ pins to construct weights $w_m \propto exp(-(\epsilon_m - \epsilon_{\min})/\sigma)$, which are used to interpolate O and so find U. Note that $\epsilon_R$ separates physically distinct stored pins but is blind to the gauge itself with every candidate within one exact Takagi family returns the same reflector image (Fig. 5d), so it can say which stored bend is nearby, never where on the gauge manifold the fibre now sits. Two retrieval baselines bracket the problem. $\hat{R}$-selected nearest pin (`rhat_nearest`) returns the stored transmission matrix of the single best-scoring pin, so the new measurement is used only to choose it, giving $F_{med} = 0.150$ with 77% of bends above $\eta \geq 0.80$; the oracle nearest pin (`oracle_nearest`), told the ground-truth answer, scores exactly the same median 0.150. The score picks the optimal library neighbour for 28 of the 30 held-out bends and the second best for the other two, so $\epsilon_R$ is close to an oracle selector.

Interpolating in the ambient matrix space does not help much. Euclidean averaging $\mathcal{N}\left(\sum_m w_m U_{j_m}\right)$ (`raw_tm_euclidean` – $\mathcal{N}$ is normalization rescaling the Frobenius norm to √n) is worse than retrieval at $F_{med} = 0.186$, because it violates the quadratic round-trip constraint $\widehat{U}^T R \widehat{U} = Q_{new}$, and multiplicative geodesic interpolation $\mathcal{N}\left(\exp\left(\sum_m w_m \log\left(U_{j_m} U_{j_1}^{-1}\right)\right) U_{j_1}\right)$ (`raw_tm_generator`) only matches it at 0.164. A physics-informed method (`physics_shape_proposal`) is the only one given information beyond $Q_{new}$: the bend coordinate itself. It weights neighbouring pins by their distance in bend radius rather than by reflector score and interpolates them multiplicatively, yet it reaches only $F_{med}$ = 0.156, no better than retrieval, even though it is handed the bend radius of the answer. (When no coordinate is available, the method instead runs a local search over trial interpolations between the two best-scoring pins, ranked by $\epsilon_R(U_{trial})$.) Knowing where the new state lies along the bend program therefore does not help if what is interpolated is the transmission matrix itself.

The gauge-interpolating methods are different in kind. Takagi Lie-gauge interpolation (`takagi_lie`) branch-aligns the gauges of the top $k = 4$ scored pins onto one sheet, interpolates only the missing gauge $\hat{O} = O_{j_1} \exp\left(\sum_{m=1}^{k} w_m \Pi_{\mathfrak{o}}\left(\log\left(O_{j_1}^{-1} O_{j_m}\right)\right)\right)$ in $\mathfrak{so}(n, \mathbb{C})$, and keeps the Takagi factor $B_{new} = D_{Q,new}^{1/2} V_{new}^T$ exactly as measured at the new configuration, so that $\widehat{U} = A_R^{-1} \hat{O} B_{new}$ is built rather than chosen. It reaches $F_{med} = 0.00228$, a $65.5 \times$ reduction over nearest-pin look-up ($68.4 \times$ if the global phase is not removed, since only the retrieval baselines pay for it). The gain is in the median: the 90th-percentile error is 0.64 for Takagi Lie-gauge against 0.85 for retrieval, and this skewness comes from the widely spaced pins at small bend radius and from wrong branches: the new Takagi factor is aligned to the best-scoring pin with signs taken from the real part of each overlap, which fails once the overlap phases have wound past 90°. Recovering the branch with the phase matching of Section 2.6 lowers the median to $5.5 \times 10^{-4}$ and the 90th percentile to 0.47 (Fig. 8d). Since the oracle scores no better than $\epsilon_R$, that gain cannot come from better pin selection: it comes from keeping $B_{new}$ exact and borrowing from the library only the one quantity a single-ended measurement cannot see, the gauge $O$.

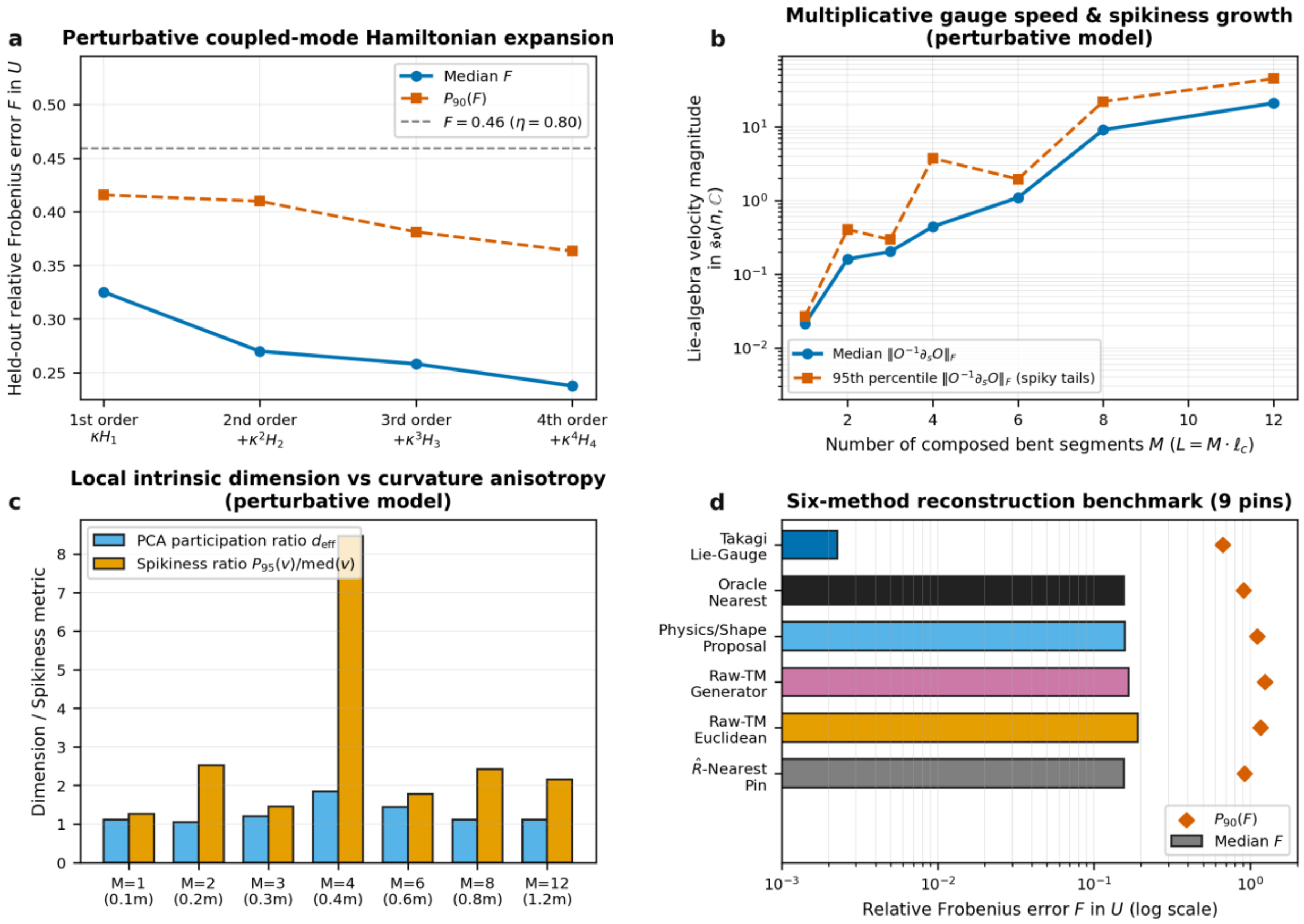

**Fig. 2.** Why the gauge is smooth on a single bend yet hard to sample across many. **a**, Fitting the conformal-mapping coupled-mode Hamiltonian $H(\kappa, \phi)$ on odd pyMMF bends and predicting the even ones improves the held-out relative Frobenius error in $U$ from $F = 0.325$ at first order ($\kappa H_1$) to $F = 0.236$ at fourth order, with the 90th percentile falling from 0.415 to 0.363. **b**, In the perturbative model, composing $M$ independent bent segments of length $\ell_c = 0.10$ m raises the Lie-algebra velocity magnitude $\| O^{-1}\, \partial_s O \|_F$ by three orders of magnitude in the program coordinate $s$, from 0.021 at $M = 1$ to 20.8 at $M = 12$, because the segment generators do not commute. **c**, The motion stays low-dimensional — the tangent-PCA participation ratio never exceeds 1.84 — but grows increasingly uneven along the path, the spikiness ratio $P_{95}(v)/med(v)$ peaking at 8.47 at $M = 4$, which is where a rapid small-radius transition is left unpinned. **d**, Six reconstruction algorithms scored identically on 30 held-out bends from 9 space-filling pins: Takagi Lie-gauge interpolation reaches a median $F = 2.28 \times 10^{-3}$, 65.5 × better than every nearest-pin scheme, which cluster at $F \approx 0.16$ however the pin is chosen. All six retain a heavy tail, $P_{90}(F) \geq 0.64$. Recovering the new state's branch by phase matching (Section 2.6, Fig. 8d) lowers these to a median F = 5.5 × $10^{-4}$ and $P_{90}$(F) = 0.47.

## 2.3 Simulated length scaling (5 cm to 5 m) and the limit of database look-up

Why do passive look-up table ("database") methods work on short or constrained laboratory setups yet fail on long, freely flexing fibres? We first established an empirical product-depth scaling baseline by multiplying $d \in \{1, \ldots, 8\}$ randomly drawn pyMMF step-index bend matrices $U_s^{(d)} = \mathcal{N}\left(A_{s,d} \cdots A_{s,1}\right)$ across library sizes $N_{pin} \in [8, 16384]$: requiring 80% of held-out product states to achieve high quality focusing ($\eta \geq 0.80 \Leftrightarrow F \leq 0.4595$) yields the exponential scaling law

$$\log_{10} N_{80\%}(d) = 0.7236 + 0.4204\, d \quad (R^2 = 0.994) \Leftrightarrow N_{80\%}(d) \simeq 13.9\,(2.63)^{d-1},$$

$$N_{50\%}(d) \simeq 5.91(1.68)^{d-1} \text{ and } N_{90\%}(d) \simeq 33.2(3.58)^{d-1}.$$

Every additional independent mixing section multiplies the required calibration library size by $\sim 2.63 \times$ ($2.61 \times$ with the global phase removed). The multiplier is $2.6 \pm 0.1$ for bends of radius 0.30 to 4.0 m and rises to 2.8–3.8 when the ill-conditioned $R = 0.20$ m bend is admitted (Fig. S2).

**Two length models.** Product depth does not say how long a fibre is: converting it to length needs a model of how a bent fibre evolves, and the answer depends on that model. We therefore used two models built in different ways. The perturbative model (Appendix A.4) chains coupled-mode transfer matrices, one per 0.10 m segment, using the bend Hamiltonian of Section 2.2. It is phenomenological: its bend couplings are random matrices shaped by a mode-group selection rule and scaled to the pyMMF mixing rate, its straight-segment phases are wrapped to 2π before being combined with the bend terms, its second-order bend terms are comparable to the first-order ones, and its mode-dependent loss, which creates the gauge, is an assumed constant. Although it uses the evenly spaced propagation constants of a parabolic graded-index core, its random, step-index-scaled couplings make it a generic phenomenological fibre rather than a model of a GRIN fibre. The mode-solver model (physical_sim_numpy.py in the code release) solves the scalar guided modes of a standard step-index fibre (core radius 25 μm, NA 0.22, 632.8 nm), builds each bent segment from the overlap integrals of those modes, propagates about 120 modes and observes the lowest 55, so that loss arises from coupling into unobserved modes rather than from an assumed term. Each model has some physical basis: the perturbative model uses the form of the bend Hamiltonian with a coupling strength scaled to pyMMF, and the mode-solver model computes its couplings from the overlaps of the guided modes. Neither has been validated against a real fibre, so we draw firm conclusions only from results on which both agree; where they differ, they indicate a range rather than a value.

Both models used the same worm-like-chain bending prior and the same bend programmes, with the global phase removed. The fibre is divided into segments of length $\ell$_c = 0.10 m, each of constant curvature; the curvatures are drawn with κ_rms = 0.80 $m^{-1}$ and correlated between neighbouring segments over a persistence length $\ell$_p = 0.16 m (correlation exp(−$\ell$_c/$\ell$_p) = 0.54). The perturbative model ran libraries of up to $10^5$ pins at seven lengths from 0.05 to 5.00 m on a GPU workstation (NVIDIA RTX PRO 6000 Blackwell, 98 GB). For a direct comparison, both models were also run at a reduced size with identical bend programmes (96 held-out fibres, libraries up to 1,024 pins; Fig. 4b,c). Coverage was evaluated at η ≥ 0.80 (F ≤ 0.4595) and at the relaxed η ≥ 0.40 (F ≤ 0.8574).

**The models differ in bend sensitivity.** For a uniformly bent 1 m fibre with R ≈ 1.5–2.3 m, the perturbative model is 1.6–2 times more bend-sensitive than pyMMF and the mode-solver model 1.2–1.7 times less (Fig. 4a). pyMMF's propagation constants differ from the exact step-index values by about 200 $m^{-1}$, so it cannot decide between them. Because covering costs grow as a power of the spread whose exponent is the effective dimension, differences of this size move the limit of look-up a long way.

**Look-up cost grows exponentially with the number of independent bends.** Both models show it. In the perturbative model, 80% coverage at $\eta \geq 0.80$ needs 8.0, 8.0 and 41.4 pins at L = 0.05, 0.10 and 0.25 m and 2,472 pins at 0.50 m (Fig. 3a); in the mode-solver model it needs about 12 pins at 1 m and 178 at 2 m, and is not reached at 5 m within 1,024 pins (Fig. 4b). The rate differs, but in both models each additional length of free fibre multiplies the library size, as the product-depth law predicts.

**There is a length beyond which look-up fails, and the models bracket its position only roughly.** In the perturbative model it lies at about 0.5 m: at 1.00 m, $10^5$ pins cover only 35% of fibres at $\eta \geq 0.80$, and 2 m and 5 m are out of reach even at $\eta \geq 0.40$ (Fig. 3a,b). In the mode-solver model it lies between 2 and 5 m: at 5 m, 1,024 pins cover 1% of fibres at $\eta \geq 0.80$ (Fig. 4b). Both models therefore place a limit within the first few metres of free fibre, but its position is set by bend sensitivity, which differs between them by about a factor of two in each direction, and it can only be fixed by measuring how quickly a real fibre decorrelates.

**Interpolating only the gauge is more accurate than look-up.** At a fixed budget of 256 pins, Takagi Lie-gauge interpolation has a held-out error one to two orders of magnitude lower than retrieval and Euclidean averaging at every length up to 1 m in both models (Fig. 4c). In the perturbative model it keeps $\eta \geq 0.80$ to L = 1.00 m (F = 0.37, against 0.78 for retrieval; Fig. 7a); in the mode-solver model at 1 m it gives F = 0.0085 against 0.27 for retrieval. Beyond 1 m the advantage narrows as all methods approach the threshold: in the mode-solver model at 5 m interpolation still reaches $\eta \geq 0.80$ for 31% of held-out fibres, against 8% for Euclidean averaging and none for retrieval, while in the perturbative model no method reaches it beyond 1 m.

**Mechanical constraint, tested in one model.** In the perturbative model, encasing straight sections of a 2.00 m fibre in rigid sleeves and restricting flexing to $J \in \{1, 2, 3, 5, 10\}$ joints reduces N_pin(L = 2.0 m) from more than $10^5$ to:

| Mechanical constraint at $L = 2.0$ m | $N_{pin}$ at $\eta \geq 0.80$ | at $\eta \geq 0.40$ |
|---|---|---|
| $J = 1$ hinge | 8.0 | 8.0 |
| $J = 2$ hinges | 45.8 | 8.0 |
| $J = 3$ hinges | 199.6 | 10.7 |
| $J = 5$ hinges | 7,527 | 98.4 |
| 3 universal joints (6 DOF) | $> 10^5$ | 1,694 |
| unconstrained spline | $> 10^5$ | $> 10^5$ |

In this model five hinges (five degrees of freedom) is the most a 2 m fibre can have before it needs more than $10^4$ pins; three universal joints (six degrees of freedom) need more than $10^5$. The mode-solver model cannot test this, because its smallest library already covers every joint configuration; the conclusion that constraint rather than length controls the cost rests on the product-depth law and one model.

**Summary.** The two models agree that look-up cost grows exponentially with the number of independent bends, that this produces a length beyond which look-up fails, and that interpolating only the gauge is more accurate than look-up. They differ on where the limit lies, about 0.5 m in the perturbative model against 2–5 m of free step-index fibre in the mode-solver model, which places it at the metre scale; fixing it more precisely needs a measurement of how a real fibre decorrelates with bend separation. These results motivate the waveguide and algorithmic routes to reducing the pin count that follow.

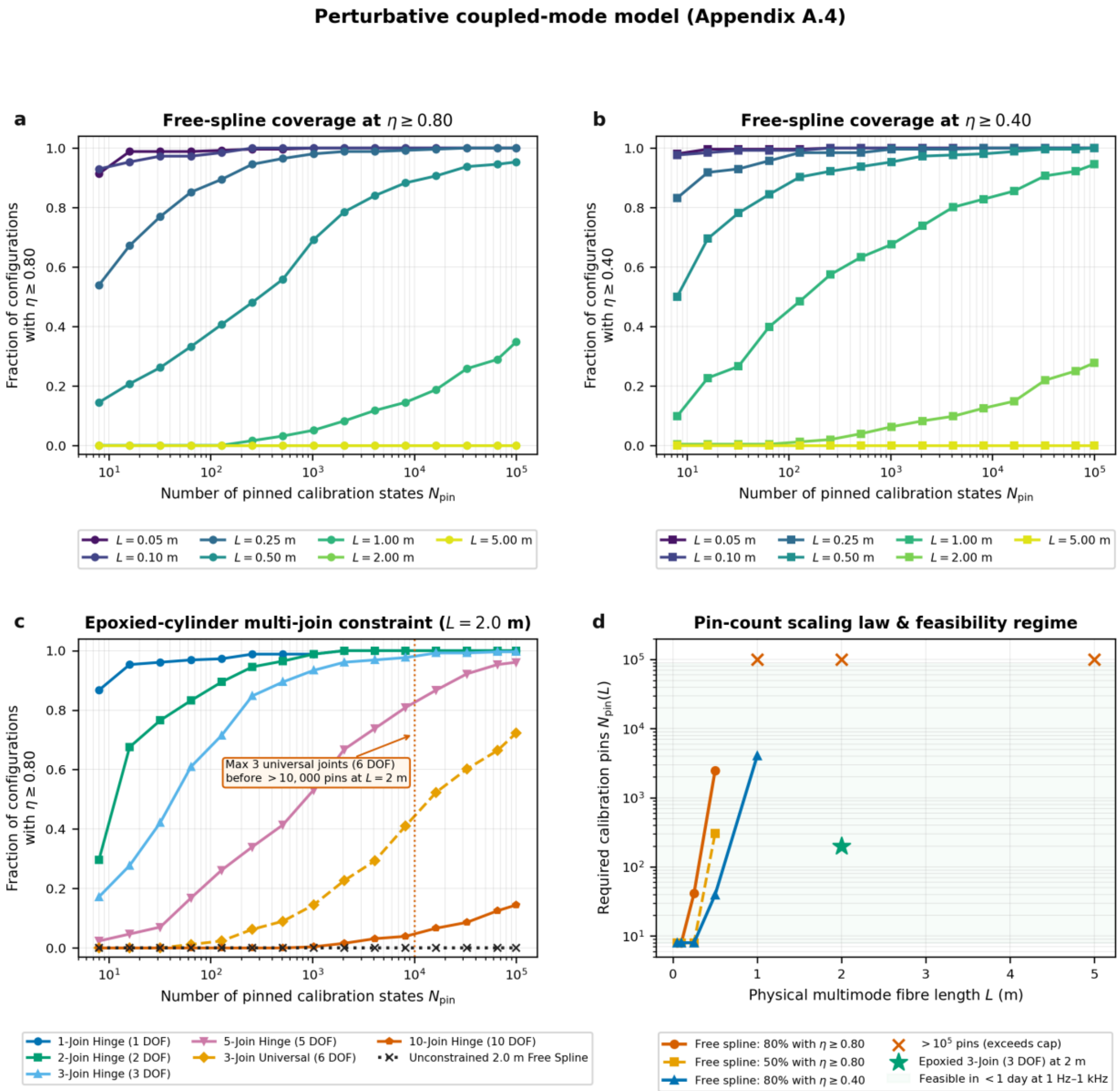

**Fig. 3.** In the perturbative model, brute-force calibration does not scale with fibre length, but mechanical constraint restores it. **a**, For a freely bending fibre (worm-like chain, $\kappa_{rms} = 0.80\ \mathrm{m}^{-1}$), the fraction of 256 random deformations whose nearest stored pin reaches $\eta \geq 0.80$, against library size. Eight pins suffice at $L = 0.10$ m and 41 at $L = 0.25$ m, but $L = 0.50$ m needs 2,472, and at $L = 1.00$ m a library of $10^5$ pins still covers only 35%; at 2 m and 5 m, nothing is covered at any budget. **b**, Relaxing the threshold to $\eta \geq 0.40$ buys roughly two decades: $L = 1.00$ m then reaches 80% coverage with 4,061 pins, while 2 m reaches only 28% at $10^5$ pins and 5 m remains out of reach. **c**, Constraining the same 2 m fibre with epoxied hinges collapses the requirement: 1 DOF needs 8 pins, 3 DOF needs 200 and 5 DOF needs 7,527, whereas 6 and 10 DOF again

exceed $10^5$ and the unconstrained 2 m spline never leaves zero. **d**, The resulting pin-count scaling law against the band reachable in under a day at 1 Hz–1 kHz; a 3-join, 3-DOF 2 m fibre sits well inside it. All curves are nearest-pin look-up without Takagi interpolation, so they bound what stored calibration alone can achieve in this model. The lengths are specific to this model; Fig. 4 compares a mode-solver model.

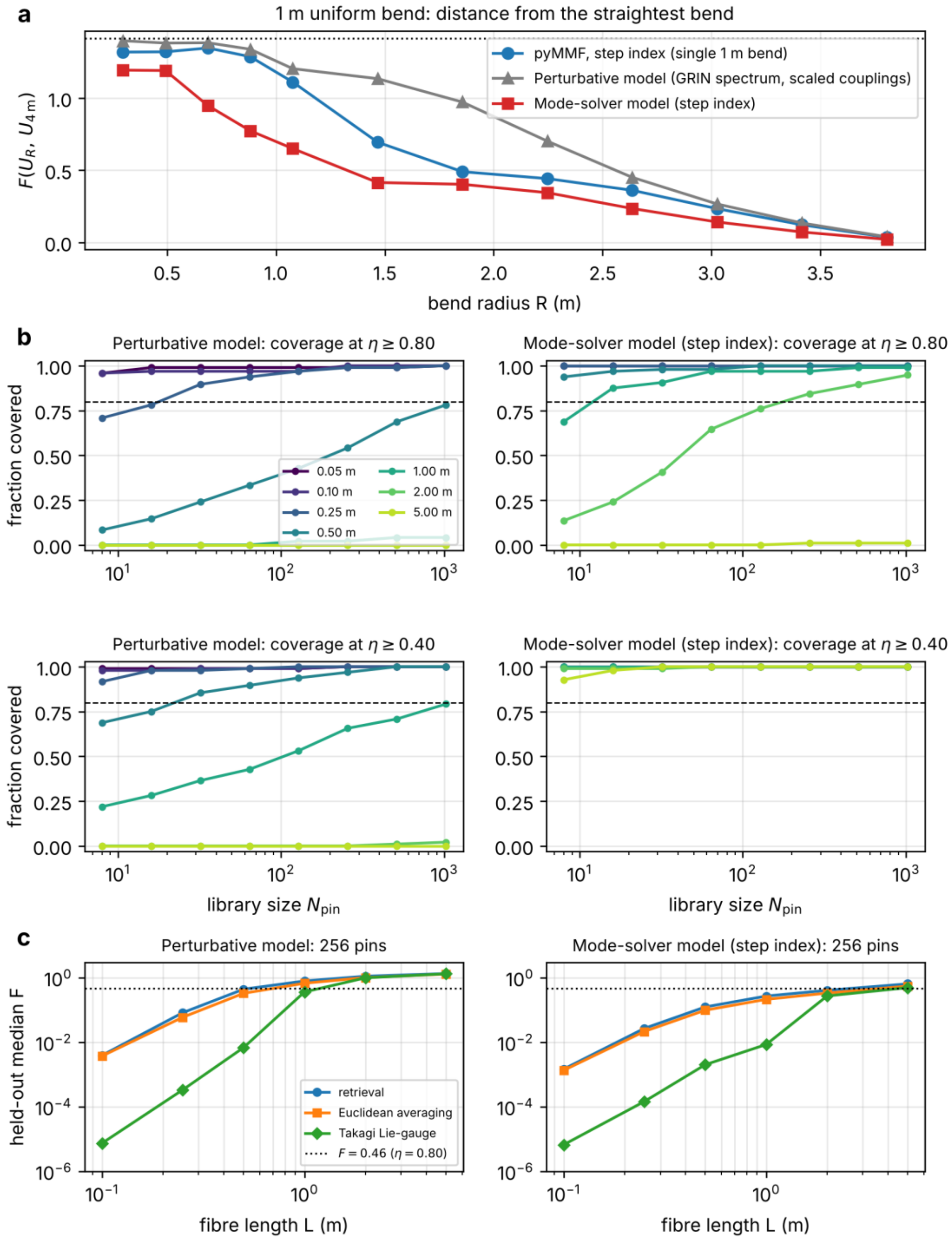

**Fig. 4.** The length-scaling results in two fibre models. **a**, Distance of a uniformly bent 1 m fibre from the same fibre bent at R = 4 m, against bend radius, for pyMMF, the perturbative model and the mode-solver step-index model, each read out on 50 modes to match pyMMF: the two length models differ in bend sensitivity by up to about a factor of two in each direction, with pyMMF between them. **b**, Fraction of 96 held-out worm-like-

chain fibres reaching η ≥ 0.80 (top) and η ≥ 0.40 (bottom) against library size, for lengths 0.05–5 m, in the perturbative model (left) and the mode-solver model (right; 55 observed modes); libraries up to 1,024 pins. The perturbative panels are a reduced re-run with the same bend programmes as the mode-solver model, so their pin counts differ from the larger GPU runs of Fig. 3. **c**, Held-out median error F at 256 pins against fibre length for retrieval, Euclidean averaging and Takagi Lie-gauge interpolation in each model; the dotted line marks F = 0.46 (η = 0.80). All distances are phase-aligned.

### 2.4 Vectorial 2N × 2N polarisation under torsional strain, structured reflectors under shot noise and three-wavelength dispersion

Conventional scalar MMF models assume an ideal circularly symmetric refractive index profile without mechanical strain, under which axial twist $\tau$ has no effect. In physical silica fibres, mechanical torsion induces shear strains $\gamma_{xz} = -\tau y, \gamma_{yz} = \tau x$ that act through the photoelastic tensor $p_{44}$ to generate Ulrich–Simon circular birefringence ($g_{elasto} = -n_0^2 p_{44} \approx 0.15$) and spin-orbit coupling ($\sigma_z \otimes L_z$), which interact with intrinsic core ellipticity ($e_{core} = 1.5 \times 10^{-3}$) and bend stress birefringence to strongly couple spatial and polarimetric modes (Appendix F).

Simulating a full $2N \times 2N$ vectorial MMF ($2N = 50$ spin-orbit modes, built on the perturbative model and sharing its mode spectrum and couplings, $L = 1.0$ m, Fig. 5a,b) across axial twists $\tau \in [-2.5, +2.5]$ rad m$^{-1}$ reveals that an average of 46.4% (and up to 73.2%) of horizontally polarised input light is scattered into the orthogonal vertical polarisation (Fig. 5a). When only a single polarisation sub-block $U_{HH} \in \mathbb{C}^{N\times N}$ is measured, this unmeasured energy loss inflates the matrix condition number to $cond(U_{HH}) = 4.08$ and causes single-polarisation Takagi Lie-gauge tracking to fail ($F_{med} = 0.697$, $\eta = 0.573$ with the global phase removed). Across twist states the retained co-polarised energy and the conditioning of $U_{HH}$ are closely related (rank correlation $-0.84$). In contrast, when both polarisations ($2N \times 2N$) are measured (Fig. 5b), unitarity is preserved ($cond(U_{2N}) = 1.01$) and vectorial Takagi Lie-gauge tracking achieves $F_{med} = 0.029$ ($\eta = 0.9991$) under twist $\tau$, a 24 × lower error than a single polarisation, and $F_{med} = 0.014$ under thermal drift $\Delta T \in$ [0,12] K (0.019 for a single polarisation; thermal drift is mostly a global phase, which the earlier phase-sensitive values of 0.053 and 0.090 included), and $F_{med} = 0.012$ across combined 2D bend-and-twist deformations.

Fig. 5d tests whether wavelength diversity can recover $O$ from $Q$ without pins. $Q$ alone does not determine $U$: the gauge $O$ must also be found, since $U = A_R^{-1}OB$. From one measurement $Q_{new}$, candidates $U_c = (W^T)^{-1} D_R^{-1/2} O_c B_{new}$ can be built from any gauge $O_c \in O(n, \mathbb{C})$; Fig. 5d uses 48, the true gauge and 47 small Lie perturbations of it. Every candidate satisfies $U_c^T R U_c = Q_{new}$ as an algebraic identity (to a relative error of 1.8 × $10^{-14}$ in the pyMMF model), so de-embedding at the central wavelength returns $\hat{R}(U_c) = U_c^{-T} Q_{new}(\lambda_0) U_c^{-1} = R$ and the same zone-plate self-image for all 48: at a single wavelength the reflector cannot tell the gauges apart, and the 48 scores agree to $10^{-8}$, so any ranking among them is arbitrary. Dispersion supplies the missing information, because it changes $U$ with wavelength in a way the gauge must follow. Recording the round trip at three wavelengths, $\lambda_0$ and $\lambda_0 \pm \Delta\lambda$ about 632.8 nm, and modelling the distal dispersion with the Lee–Villiger second-order parametric model[27], breaks the degeneracy: a wrong gauge predicts the wrong $Q$ at $\lambda_0 \pm \Delta\lambda$, and for every span of 0.1 nm or more the true gauge ranks first of 48, at $F = 1.7 \times 10^{-8}$.

Once pins exist, the ambiguity is broken and one wavelength suffices (**Fig. 5c**). With $\hat{O}$ constrained by Lie interpolation across neighbouring pins — or with local trials proposed from $(O, B)$ without rebuilding from $B_{new}$ , in order to find out which reflector gives the most accurate reconstruction of U under realistic noise conditions, we simulated shot noise with 1.5 $e^-$ camera read noise across input budgets $N_{in} \in [10^3, 10^6]$ photons per input mode. A bare 4% Fresnel facet (`uniform_fresnel`) fails at $10^4$ photons per mode ($\eta = 0.269$, $F = 0.981$), and has 39 near-repeated Takagi values ($cond(R) = 1.30$). When Takagi values coincide, the corresponding Takagi vectors are determined only up to a real orthogonal rotation within each degenerate block. The measured factor $B_{new} = VD_Q^{1/2}$, and hence the reconstruction Û can therefore be wrong by an unknown rotation, and because every rotated V reproduces Q equally well, no residual on the measurement can detect it.

Photon starvation is not the cause. Pixel reflectivity is not photon return: the 85% random-phase mirror returns only 5.6% of the light into the guided modes (mode-solver model at 660 nm, where bare glass returns 3.6%). The mirror is highly reflective and yet the random phases diffract the light through steep angles outside what the fibre can guide – so much of the return signal is lost. Compare the random phase mirror to 3.6% for bare glass; its noise advantage is √(5.6/3.6) ≈ 1.25×, not the √(85/4) ≈ 4.6× its reflectivity suggests. Even with no noise the flat facet stops at $F = 0.12$–$0.20$ while every structured reflector reaches $10^{-3}$ or better, and a $2 \times 2$ factorial design of spectrum against photon budget shows the spectrum dominates: giving the flat facet $21 \times$ more light moves it only from $F \approx 0.99$ to 0.86–0.90, still worse than a dim structured reflector (0.79). The design rule is to break the reflector's Takagi spectrum, not to reflect more light. A 4%-floored FZP (`fzp_4pct_floor`, 42% mean reflectivity, $cond(R) = 3.40$), an ideal binary FZP (40%), a 4%-floored random binary mask (47%) and a metasurface FZP (55%) all stay above threshold at the same budget, at $\eta = 0.865$–$0.915$.

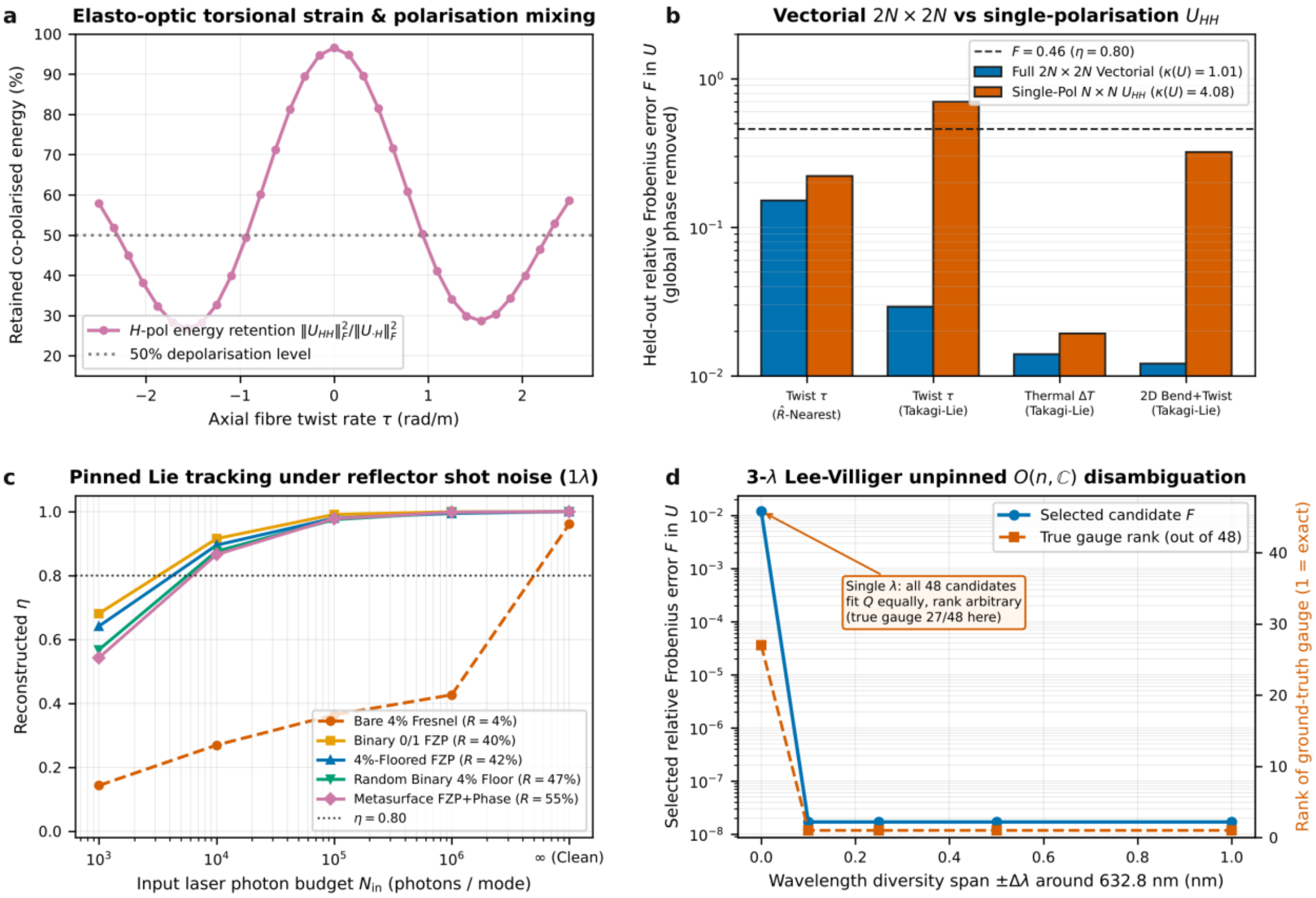

**Fig. 5.** Vectorial treatment, reflector photon budget and wavelength diversity. **a**, Axial twist mixes polarisation strongly: the co-polarised energy fraction $\| U_{HH} \|_F^2 / \| U_{\cdot H} \|_F^2$ falls from 97% at $\tau = 0$ to 27% near $\tau = \pm 1.6$ rad m$^{-1}$, crossing the 50% level within 1 rad m$^{-1}$. **b**, Reconstructing the full $2N \times 2N$ vectorial matrix rather than the single-polarisation block $U_{HH}$ matters most where that mixing is strongest: under twist, Takagi Lie-gauge interpolation reaches $F = 0.029$ vectorially against $F = 0.697$ for $U_{HH}$ , the only case in the figure that fails $\eta \geq 0.80$ , and the advantage persists under thermal drift (0.014 against 0.019) and combined 2D bend and twist (0.012 against 0.321), following the conditioning $\kappa(U) = 1.01$ against 4.08. **c**, Under Poisson shot noise at the distal reflector, a bare 4% Fresnel facet never reaches $\eta = 0.80$ at any budget tested, while all four structured reflectors ($R = 40$–55%) pass it at $10^4$ photons per mode and exceed $\eta = 0.99$ by $10^6$. **d**, At a single wavelength all 48 candidate gauges reproduce $Q$ equally well (scores agree to $10^{-8}$), so their ranking is arbitrary (here the true gauge falls 27th and the selected candidate has $F = 0.012$); a span of $\pm 0.10$ nm around 632.8 nm ranks the true gauge first and drives the selected error to $2 \times 10^{-8}$.

## 2.5 Regularised reflector scoring on measured 5 m bent fibres

We tested reflector de-embedding using 139 measured transmission matrices from a 5.0 m, 50 µm-core step-index multimode fibre at 660 nm[29]. The dataset also contains measured intensity reflection matrices and 100-point overhead-camera splines for 137 fibre configurations. The closest recorded shapes are approximately 1.3 cm apart (median pairwise separation 6.31 cm). The nearest measured neighbour of each TM differs from it by a median F = 0.94 (η ≈ 0.31) and by at least F = 0.67 (η ≈ 0.61), against F = 0.46 for η = 0.80, so the available TMs do not provide a sufficiently close calibration neighbourhood to demonstrate interpolation of the aligned gauge manifold $\mathcal{O}$. We therefore used the dataset to test whether reflector de-embedding could identify calibrations optically close to a held-out target.

The measured TMs were projected onto a fixed 145-mode basis. We simulated an eight-zone Fresnel zone plate with amplitude reflectivities of 0.9 in its open zones and 0.2 elsewhere, corresponding to power reflectivities of 81% and 4%. Its known modal reflection matrix $R$ was used to generate a synthetic round trip for each target:

$$Q_{\text{target}} = {U_{\text{target}}}^{\mathrm{T}} R U_{\text{target}}$$

Each target TM was then excluded from the library, leaving 138 measured calibration candidates. For a candidate $U_j$, conventional de-embedding estimates the reflector as

$$\widehat{R}_j = U_j^{-\mathrm{T}} Q_{\text{target}} U_j^{-1}$$

The measured modal TMs are poorly conditioned, so applying their inverses twice can amplify weak transmission directions and distort the candidate scores. We therefore replaced each inverse with its Tikhonov-regularised form,

$$A_j = \left(U_j^{\mathrm{H}} U_j + \lambda s_{1,j}^2 I\right)^{-1} U_j^{\mathrm{H}}$$

where $s_{1,j}$ is the largest singular value of $U_j$, and calculated

$$\widehat{R}_j = A_j^{\mathrm{T}} Q_{\text{target}} A_j$$

Candidates were ranked by the relative Frobenius residual between $\widehat{R}_j$ and the known $R$, after global-phase alignment. Regularisation suppresses weak transmission directions, stabilising the score when the exact target is absent.

With no added noise on $Q$, $\lambda = 0.1$ identified the true nearest remaining TM in 117/139 cases (84.2%), compared with 4/139 (2.9%) using the exact inverse (Fig. 6). When the regularisation parameter was fixed using 28 development states, success on the remaining 111 targets was 83.8%. These results demonstrate held-out nearest-calibration retrieval using measured TMs and simulated complex round trips; they do not establish reconstruction of the held-out TM.

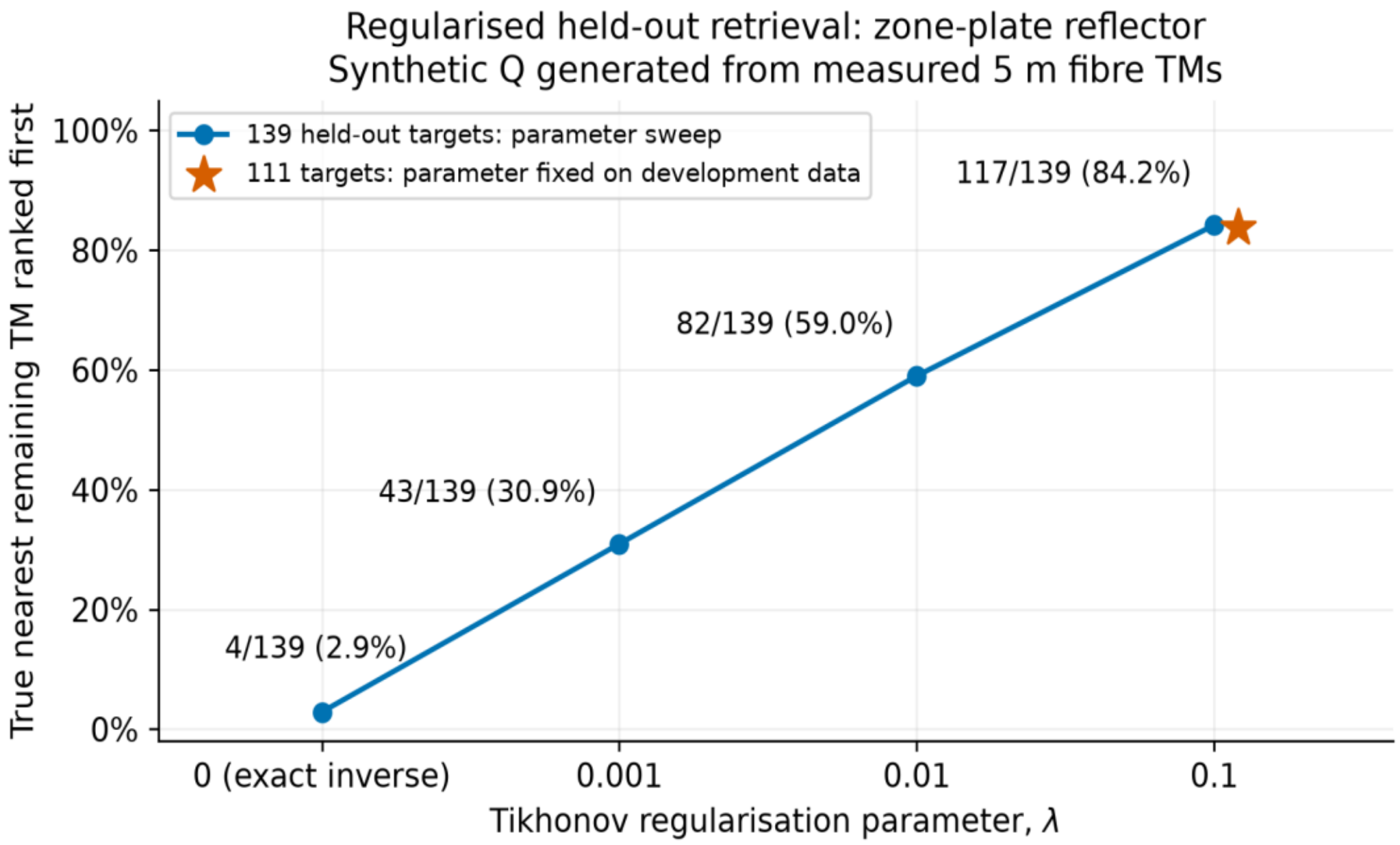

**Fig. 6.** Regularised held-out reflector scoring. Synthetic Q is generated from measured modal TMs using the known zone-plate reflector, without added noise. Each test TM is absent from its 138-candidate library. The curve shows nearest-TM identification across 139 targets as the regularisation parameter increases; counts label each point. The star shows 93/111 correct identifications (83.8%) with the parameter fixed using 28 development states. Success means selecting the nearest available measured TM, not reconstructing the held-out TM.

## 2.6 Multi-length solver, 500-mode GPU stress test and single-ended branch recovery

To test how the method scales with fibre length and mode count, we ran Takagi Lie-gauge reconstruction in the perturbative model (Fig. 7). At a fixed budget of $N_{\text{pin}} = 256$ across L ∈ {0.10, 0.25, 0.50, 1.00} m, it reduces the held-out relative Frobenius error in U by one to two orders of magnitude relative to rhat_nearest and raw_tm_euclidean, and in this model keeps η ≥ 0.80 out to L = 1.00 m (F = 0.3693 < 0.4595, η = 0.8682) with those 256 pins alone (Fig. 7a); the same comparison in the mode-solver model is shown in Fig. 4c.

Mode count is not the obstacle it might appear. In the perturbative model, for a 500-mode waveguide — 500 × 500 complex128 matrices, 124,750 complex parameters in $\mathfrak{so}(500, \mathbb{C})$ — a single Autonne–Takagi factorisation executes in 47.2 ms on an NVIDIA RTX PRO 6000 Blackwell GPU with relative residual $3.68 \times 10^{-12}$, reaching, in the perturbative model, 95.3% coverage at $\eta \geq 0.80$ with 256 pins and 100% coverage ($F_{med} = 0.0201$) at 4,096 (Fig. 7b).

Single-ended branch recovery. The held-out results above, and those of Section 2.1, take the Takagi branch of each withheld state from its own transmission matrix. A live measurement does not supply it: its Takagi factor $V_{\text{live}}$ arrives on an unknown one of the $n!2^n$ sheets, and Q, its Takagi values and the de-embedded reflector $\hat{R}$ are identical on every sheet, so the branch can only be inferred by continuity from the calibration pins. The pipeline of Fig. 2d aligns $V_{live}$ to the best-scoring pin by Hungarian assignment on $|V_{pin}^H V_{live}|$, with signs from the real part of each matched overlap. That fails once the overlap phases have wound past 90°, and they wind fast: in the mode-solver model of a uniformly bent 1m step-index fibre (50 modes, $\kappa \in [0.25, 3.33]$ $m^{-1}$), between two neighbouring pins of 16 evenly spaced in curvature, the overlap phase of the fastest Takagi vector turns through 713° while its magnitude stays above 0.80 (Fig. 8b). The Fig. 2d alignment therefore takes a wrong branch that raises $F$ by more than 0.05 above its value with the true branch (defined as a harmful branch error, so that wrong branches that barely change the reconstruction are not counted) for 68%, 47% and 30% of live states with 16, 32 and 48 pins (Fig. 8a).

We recover the branch and the position from Q, R and the pins alone, in four steps. First, the reflector residual $\varepsilon_j = \|U_j^{-T} Q U_j^{-1} - R\|_F / \|R\|_F$ is computed at every pin, and the vertex of a symmetric V through the lowest score and its two neighbours estimates where the live state lies between them (Fig. 8c). Second, calibration records, for every Takagi vector, the phase drift of its overlap from one pin to the next. Because the phase can turn by more than $2\pi$ between pins, the drift is known only modulo $2\pi$, so the phases are unwrapped by choosing the multiple of $2\pi$ in each interval that makes the phase rate in curvature as smooth as possible (a dynamic programme anchored at the gentlest bend). The live state's position between the pins is then found from its overlap phases. For each Takagi vector, the phase of the live state's overlap with the best pin is measured. It is known only modulo $\pi$, because the sign of each live vector is unknown. For any trial position between the neighbouring pins, the calibration table predicts what each of these phases should be. A single vector cannot fix the position: its phase returns to the same value modulo $\pi$ many times between pins, once for every half-turn it makes. But the vectors turn at different rates, so their sets of possible positions are spaced differently, and only at the true position do all of them agree. The position is taken where the mismatch, summed over all vectors and weighted lightly towards the $\hat{R}$ estimate of the first step, is smallest (Fig. 8c). Third, the columns are assigned by Hungarian matching on $1 - |C_{ij}| + 5|\hat{d}_i - d_j|/\max \hat{d}$, where $\hat{d}$ are the Takagi values interpolated to that position, and each sign is taken against the predicted phase. Fourth, the gauge is PCHIP-interpolated in $\mathfrak{so}(n, \mathbb{C})$ at that position, exactly as in Section 2.1. Each sign decision also yields a confidence $S = \min_k |\mathrm{Re}\, z_k|/|z_k|$, where $z_k$ is the matched overlap rotated back by its predicted phase, and states with $S < 0.8$ are flagged. The same code accepts a measured curvature in place of the reflector-residual location $\varepsilon_j$ of the first step when the fibre shape is known; the phase matching then refines it as before.

**Table 1.** Branch recovery for held-out states under four protocols. Oracle: the branch from the held-out state's own transmission matrix. Curvature known: steps two to four at the true curvature. Single-ended: all four steps from Q, R and the pins. Fig. 2d alignment: best-$\hat{R}$ pin, signs from the real part of the overlaps, $\hat{R}$-weighted gauge. pyMMF columns give the median phase-aligned F followed by the number of held-out bends with $\eta < 0.80$ (second column) or the 90th-percentile F (third column); model columns give harmful branch errors (%) with pins evenly spaced in curvature. Noise-free.

| Protocol | pyMMF, every second bend withheld: median F; bends with $\eta < 0.80$ (of 19) | pyMMF, 30 bends from the 9 pins of Fig. 2d: median F; $P_{90}(F)$ | Mode-solver, 16 / 32 / 48 pins: harmful % | Perturbative, 16 / 32 / 48 pins: harmful % |
|---|---|---|---|---|
| Oracle | $2.66 \times 10^{-4}$; 0 | $3.24 \times 10^{-4}$; 0.038 | reference | reference |

| Protocol | pyMMF, every second bend withheld: median F; bends with η < 0.80 (of 19) | pyMMF, 30 bends from the 9 pins of Fig. 2d: median F; $P_{90}$(F) | Mode-solver, 16 / 32 / 48 pins: harmful % | Perturbative, 16 / 32 / 48 pins: harmful % |
|---|---|---|---|---|
| Curvature known | $2.66 \times 10^{-4}$; 1 | $3.24 \times 10^{-4}$; 0.40 | 0 / 0 / 0 | 26.5 / 0.3 / 0 |
| Single-ended | $3.95 \times 10^{-4}$; 1 | $5.53 \times 10^{-4}$; 0.47 | 0.3 / 0.3 / 0.3 | 25.6 / 0.3 / 0 |
| Fig. 2d alignment | $8.12 \times 10^{-4}$; 2 | $2.28 \times 10^{-3}$; 0.64 | 68.4 / 47.2 / 29.6 | 64.1 / 3.3 / 0 |

Table 1 summarises the comparison. In the mode-solver model the single-ended pipeline leaves 0.3% harmful branch errors with 16 to 48 evenly spaced pins and none with 64, against none when the curvature is known; with pins displaced at random by up to 0.4 of the spacing it leaves 1.2–2.9% from 24 to 64 pins, where the known curvature leaves none, and both leave 15% at 16 displaced pins (Fig. 8a). The perturbative model, whose phases wind fastest, needs more pins: 25.6–29.1% harmful at 16 pins, as with the curvature known, falling to 0.2–0.3% at 32 and to none at 48. The two core-profile models of Section 2.7 (GRIN and step index), where even the Fig. 2d alignment fails for at most 9.7% of states, stay at or below 0.4% at every density. Adding complex-symmetric noise of relative size $10^{-4}$ to every measured Q, pins and live states alike, leaves these rates unchanged at 32 pins. The $\hat{R}$ valley alone places the live state to a median 0.040 of the local pin spacing, and phase matching refines this to 0.011 (Fig. S3a); on the pyMMF bends, with every second bend withheld, the final error is 4.4 mm in bend radius. With every second pyMMF bend withheld the single-ended median is $3.95 \times 10^{-4}$, against $2.66 \times 10^{-4}$ with oracle alignment, and only the tightest withheld bend (R = 0.39 m) falls below η = 0.80, as it does with the curvature known (Fig. S3b). With the nine pins of Fig. 2d the median falls from $2.28 \times 10^{-3}$ to $5.5 \times 10^{-4}$ and the 90th percentile from 0.64 to 0.47 (Fig. 8d); the bends that still fail are the three tightest (R ≤ 0.59 m), where the first two pins are 0.49 m apart in radius but 2.1 $m^{-1}$ apart in curvature.

The sign confidence flags the failures that matter most. In the perturbative model with 16 pins it flags every harmful state (79 of 79 with even pins, 270 of 270 with displaced pins) at the cost of accepting only 41–43% of states, and at 32 pins it accepts 78–82% with no harmful state among them. The flag is weaker where the calibration's phase prediction is itself wrong, as happens when the unwrapping between widely spaced pins picks the wrong number of turns: the sign is then wrong but agrees confidently with the prediction, even at the true curvature. With 16 displaced pins in the mode-solver model the flag catches 88 of the 147 harmful states (the other 59 are equally harmful with the curvature known), and on the Fig. 2d pins it flags the three failing bends but accepts one with $F$ = 0.28.

Known-shape mode. When the shape is measured independently, by a shape sensor or an instrumented robot, its curvature can replace the $\hat{R}$ step. Used directly, a sensor reading needs an error below about 0.03 of the pin spacing: at 0.1 the harmful rate in the mode-solver model rises to 1.6–28.5%. Passed through phase matching, a reading within 0.1 of the spacing does as well as the single-ended pipeline (0–0.3% harmful), and one within 0.3 stays below 5%. A reading a whole spacing off is worse than using no sensor (34–38% against 0.3–2.4%), because the search then starts between the wrong pins (Fig. S4). Phase matching thus turns a coarse shape estimate into a precise one, though the single-ended pipeline does not need one.

Single-ended operation is possible only if states with close $Q$ have close $U$. Across every pair of states in the four single-bend sets, no pair with $d_Q < 0.2$ has $d_U > 0.14$, and the eight-bend shape pairs of Section 2.7 follow the same trend ($d_U/d_Q$ = 0.65 for GRIN and 0.75 for step index; Fig. S5).

What limits the pin density for single-ended use is the sampling of the overlap phases: neighbouring pins must be close enough that each phase can be unwrapped from one interval to the next and is close to linear within each interval. Two further limits remain. Even with the true branch and curvature, the interpolated gauge reaches $F \approx 0.3$ at a few states in every model (99th percentile up to 0.27 in the mode-solver model). In the perturbative and GRIN models most of these states lie where two Takagi values nearly cross; in the mode-solver model most do not, and their cause is open. And these tests use single uniform bends; multi-segment fibres, whose gauge velocity is far more uneven (Fig. 2b,c), remain to be tested.

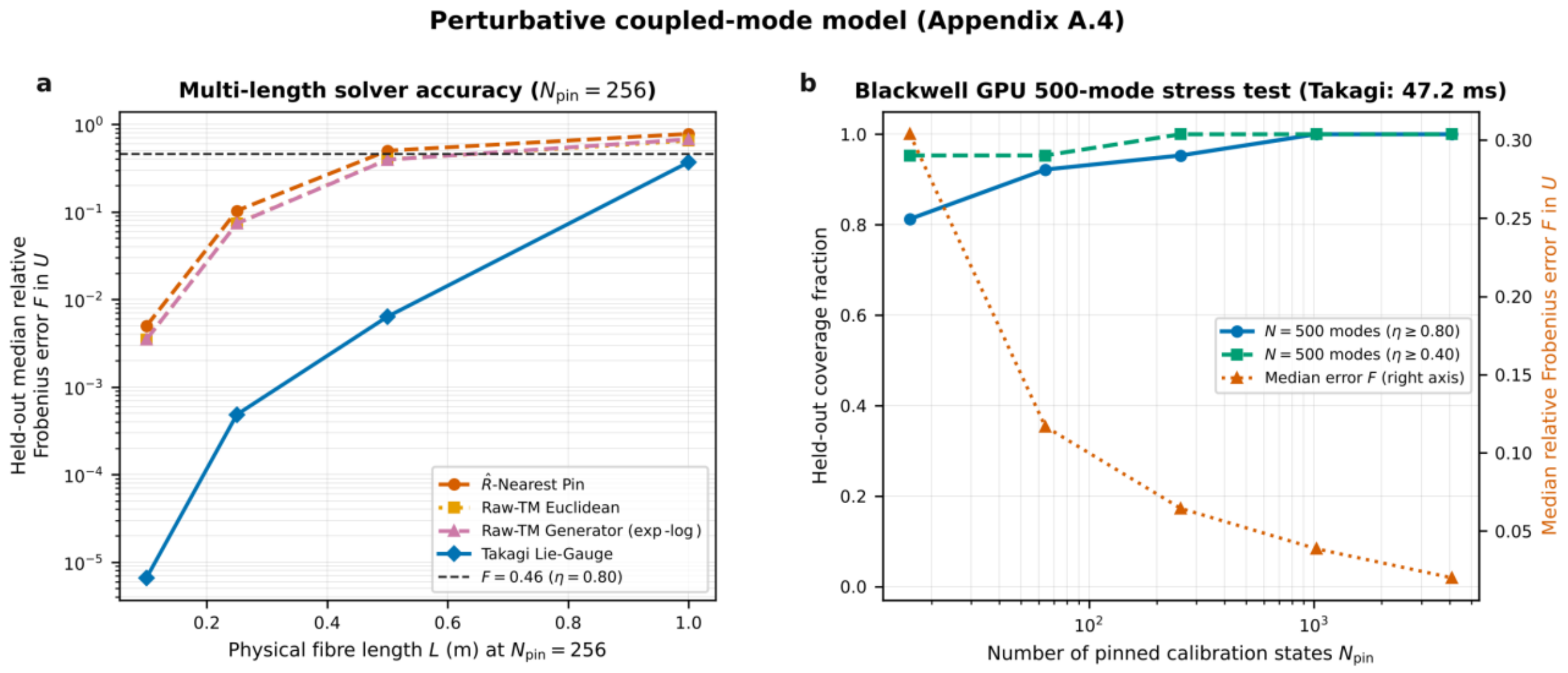

**Fig. 7.** Lie-gauge interpolation and mode-count scaling. **a**, In the perturbative model, with a fixed budget of 256 pins, Takagi Lie-gauge interpolation holds the held-out error below the usable threshold out to $L = 1.00$ m ($F = 0.369$), where $\hat{R}$-nearest-pin look-up (0.779), and raw-TM Euclidean (0.652) have both failed; at $L =$ 0.10 m it reaches $F = 7 \times 10^{-6}$, two to three orders of magnitude below any alternative. **b**, The approach is not confined to small mode counts: a single $500 \times 500$ Autonne–Takagi factorisation takes 47 ms on an RTX PRO 6000 with a residual of $3.7 \times 10^{-12}$, and 500-mode held-out coverage in the perturbative model at L = 0.25 m rises from 81% at 16 pins to 100% by 1,024.

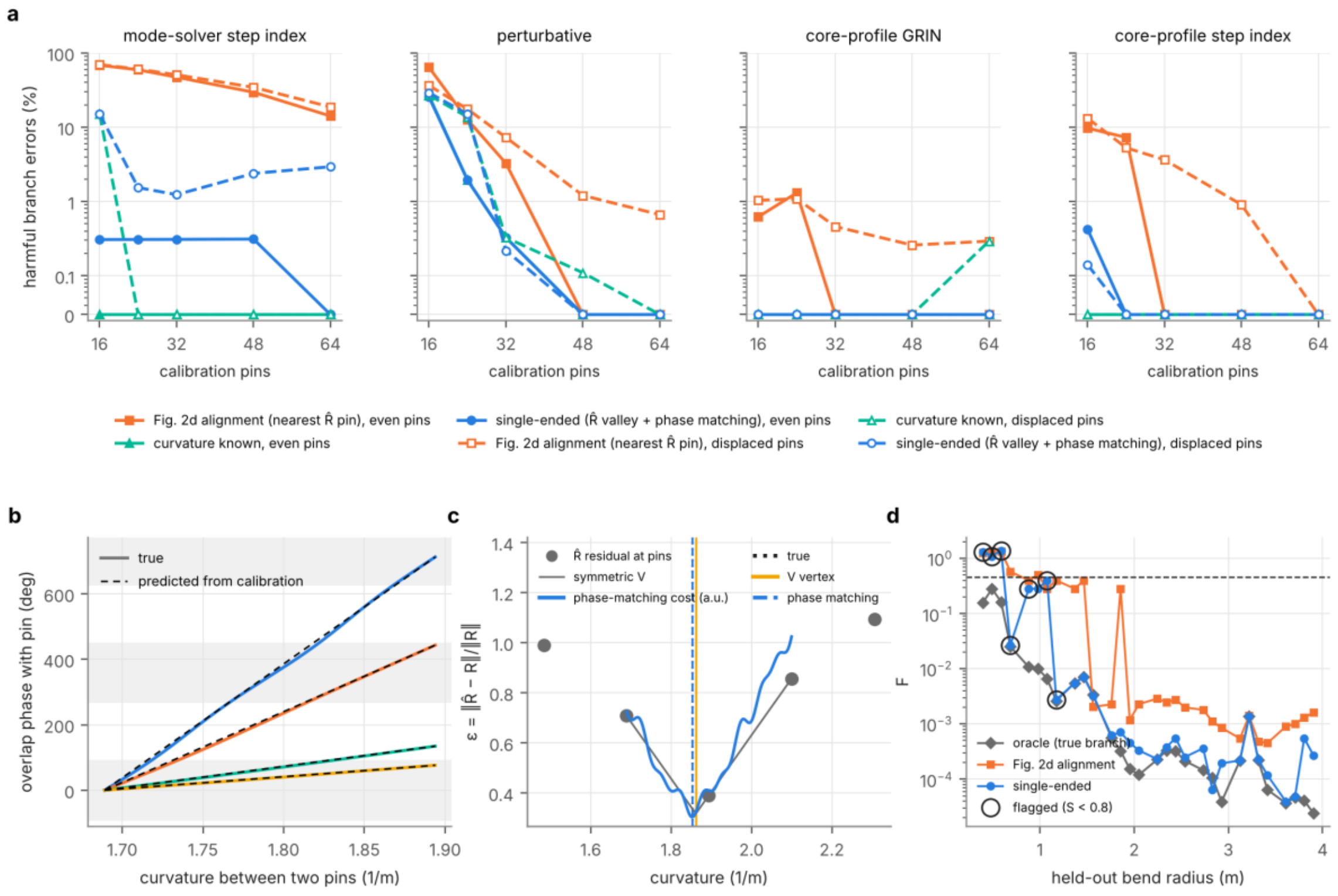

**Fig. 8.** Single-ended branch recovery. **a**, Harmful branch errors (a wrong Takagi branch raising the reconstruction error by more than 0.05) against the number of calibration pins in four fibre models, for the Fig. 2d alignment (best $\hat{R}$ pin, signs from the real part of the overlaps), the single-ended pipeline ($\hat{R}$ valley, phase matching, Takagi-value continuity) and the same branch step given the true curvature. Filled markers and solid lines, pins evenly spaced in curvature; open markers and dashed lines, pins displaced at random by up to 0.4 of the spacing (three pin sets). The lowest row marks zero. **b**, Overlap phase of the four fastest-winding Takagi vectors with a pin, between two neighbouring pins of 16 in the mode-solver model: solid, true; black dashed, predicted from the unwrapped calibration drift; shaded, where the sign from the real part of the overlap is correct. **c**, Locating one live state: the reflector residual at the pins (grey dots), the symmetric V through the lowest residual and its two neighbours (grey lines) and its vertex (yellow line), and the phase-matching cost from the overlap phases (blue), whose minimum (dashed) lies 0.002 of the pin spacing from the true position (dotted), against 0.043 for the valley vertex. **d**, The 30 held-out pyMMF bends with the 9 pins of Fig. 2d: oracle alignment, the Fig. 2d alignment and the single-ended pipeline; circles mark states flagged by sign confidence $S < 0.8$; dashed line, $\eta = 0.80$. Noise-free; phase-aligned F.

## 2.7 GRIN versus step-index manifold smoothness, quadratic Lie-algebra closure and shape-Hamiltonian degeneracy

Our multi-segment composition results (Fig. 2b,c and Appendix D.2) raise a waveguide-physics question: does compounding multiple bends make every multimode fibre manifold exponentially spiky, or does a Graded-Index (GRIN) fibre preserve a smooth, low-dimensional Takagi manifold because its bend Hamiltonian belongs to a space of quadratic oscillator operators regardless of the number of bends? In short is there a refractive-index profile that makes the gauge manifold as smooth and easy to interpolate as possible under bending?

To test this hypothesis and compare core refractive-index profiles $n(r)$, we built an exact 96-point Gauss–Laguerre spectral radial Helmholtz solver,

(experiments/e7_grin_vs_step_index_and_optimal_core.py)

parameterising the core profile over power-law exponent $\alpha \in [1,50]$ ($\alpha = 2$ parabolic GRIN, $\alpha \to \infty$ step-index) and fourth-order non-paraxial aberration coefficient $\beta_4$ (Fig. 9):

$$n^2(r) = n_{core}^2 \left[1 - 2\Delta\left((1-\beta_4)\left(\frac{r}{a}\right)^{\alpha} + \beta_4\left(\frac{r}{a}\right)^4\right)\right].$$

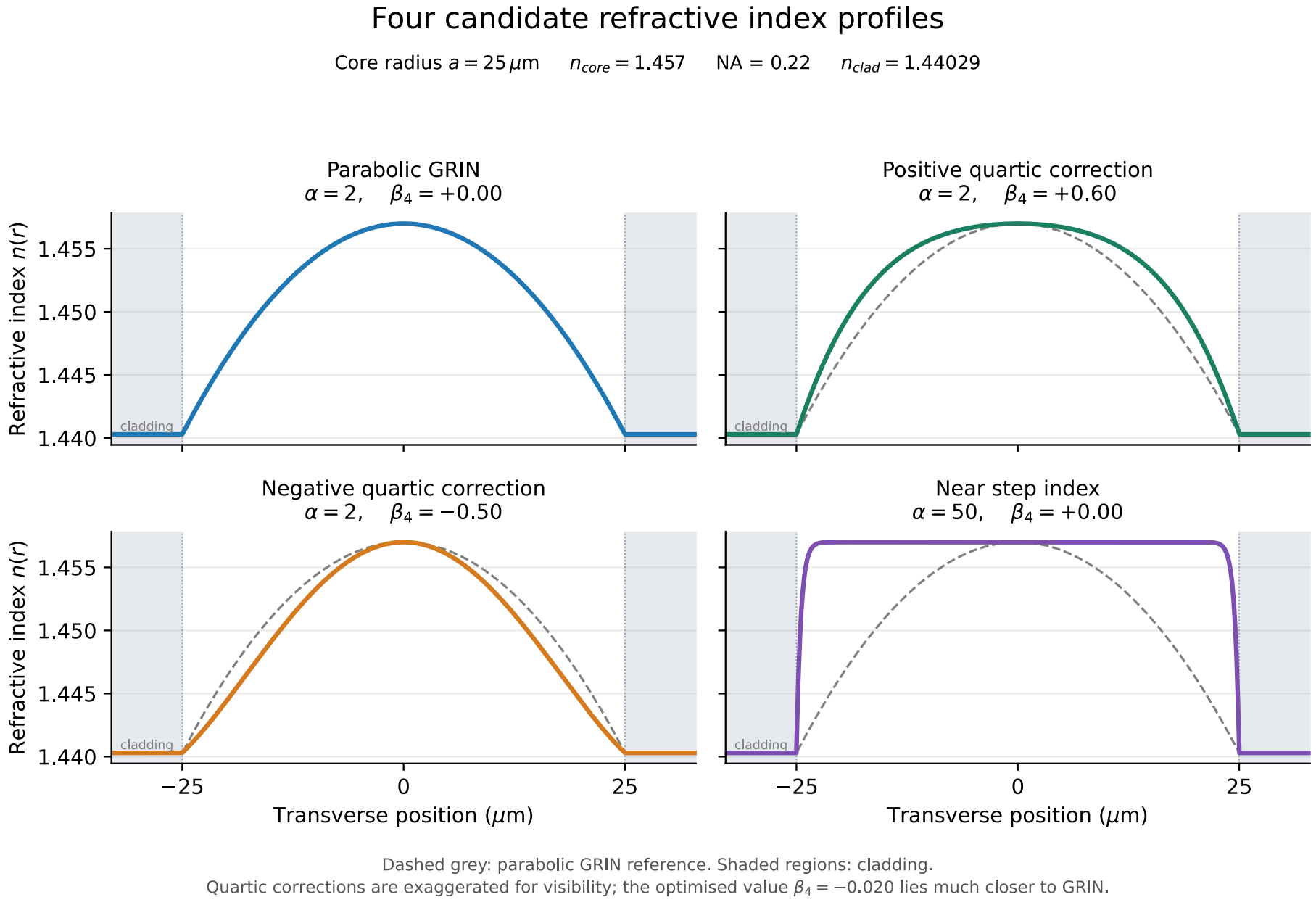

**Fig. 9.** The core-profile family searched in Section 2.7, $n^2(r) = n_{\text{core}}^2\left[1 - 2\Delta\left((1-\beta_4)(r/a)^{\alpha} + \beta_4(r/a)^4\right)\right]$, shown for $a = 25$ μm, $n_{\text{core}} = 1.457$ and NA = 0.22 ($n_{\text{clad}} = 1.44029$): parabolic GRIN ($\alpha = 2$, $\beta_4 = 0$), positive and negative quartic corrections ($\alpha = 2$, $\beta_4 = +0.60$ and $-0.50$, exaggerated for visibility) and near step index ($\alpha = 50$). Dashed grey, parabolic reference; shaded, cladding.

All results in this section come from the spectral core-profile solver with its own bend sampler and bend strengths: they compare core profiles within that solver and are not directly comparable with the two length models of Section 2.3. Within this solver we find four results (**Fig. 10a–d**). Consider first the Lie-algebra closure that a parabolic core enforces (**Fig. 10a**). In an ideal 2D parabolic harmonic oscillator ($\alpha = 2$), the unperturbed Hamiltonian $H_0 = diag(\boldsymbol{\beta}_0)$ has equally spaced principal mode groups ($\beta_g = \beta_0 - g\Omega_0$, $g = 2p + |\ell| + 1$), while the dipole bend perturbation $\kappa_x(z)C_x + \kappa_y(z)C_y$ is linear in the transverse coordinates $(x, y)$ and obeys the strict selection rule $|\Delta g| = 1$. Consequently, the second nested commutator closes back onto $C_x$ and $C_y$ identically:

$$[H_0, [H_0, C_x]] = \Omega_0^2 C_x$$
$$[H_0, [H_0, C_y]] = \Omega_0^2 C_y$$

Each bend-coupling matrix, after taking the commutator with $H_0$ twice, returns to itself multiplied by the square of the mode-group spacing, $\Omega_0^2$. This identity holds exactly for the ideal parabolic harmonic-oscillator model.

Across the sweep $\alpha \in [1,50]$ (Fig. 10a), the relative oscillator double-commutator leakage:

$$\| [H_0, [H_0, C_x]] - \Omega_0^2 C_x \|_F / \| [H_0, [H_0, C_x]] \|_F$$

drops from 0.3537 at $\alpha = 1.0$ and 0.6740 at $\alpha = 50$ (approx. Step Index), down to a minimum of 0.0013 at the parabolic profile ($\alpha = 2.00$; the residual comes from the non-paraxial propagation constants and vanishes in the paraxial limit), where the participation-ratio dimension of the 8-order nested Krylov commutator sequence $\{ad_{H_0}^k(C_x)\}_{k=0}^7$ collapses to $\dim_{Krylov} = 2.000$ (vs 3.80–4.13 for non-parabolic / step-index profiles).

Thus, regardless of how many bends $M$ are compounded in series, the Baker–Campbell–Hausdorff series of a perfect parabolic GRIN fibre remains trapped inside the finite-dimensional inhomogeneous symplectic / oscillator Lie subalgebra $\mathfrak{isp}(4, \mathbb{R})$, whereas in a step-index fibre the hard core-cladding wall breaks equal mode-group spacing and causes nested commutators to leak across the full $n^2$-dimensional algebra $\mathfrak{u}(n)$. That closure has a direct geometric consequence (Fig. 10b).

Because the interaction-picture bend operator $e^{-izH_0} C_{x,y} e^{izH_0} = \cos(\Omega_0 z) C_{x,y} - i\sin(\Omega_0 z)\Omega_0^{-1}[H_0, C_{x,y}]$ oscillates at the single pitch frequency $\Omega_0$, the end-to-end effective Hamiltonian depends to leading order only on the resonant Fourier integrals $\int_0^L \kappa_{x,y}(z) e^{\pm i\Omega_0 z}\, dz$. Any shape perturbation orthogonal to $\{1, \cos(\Omega_0 z), \sin(\Omega_0 z)\}$ is therefore degenerate in $H_{\text{eff}}$. Testing pairs of eight-bend trajectories differing by 84.8% RMS curvature in the pitch-orthogonal subspace, the gauge divergence $d_O = \| O^{(B)} - O^{(A)} \|_F / \sqrt{n}$ is five times smaller in GRIN ($d_O = 0.0483$–$0.0516$ at $\alpha \in [1.97, 2.00]$, $d_U = 0.0406$) than in step index ($d_O = 0.2467$, $d_U = 0.6120$): many distinct macroscopic 3D shapes share essentially the same effective Hamiltonian and the same gauge. The transmission-matrix degeneracy is narrow in the core exponent: $d_U$ stays below 0.05 only for $\alpha = 1.97$–$2.05$ and is at least 0.22 elsewhere (Fig. 10b), so the effect needs a near-ideal parabolic profile; the gauge divergence alone stays below 0.07 for every $\alpha \leq 2$.

The simulated parabolic GRIN fibre exhibited a more slowly varying Takagi gauge than the near-step-index fibre under the tested multi-bend deformations. Along the sampled eight-bend trajectory, its median gauge speed was 2.63, compared with 5.93 for near step index, and its 95th-percentile gauge speed was 12.55, compared with 23.37. These correspond to approximately 2.3-fold and 1.9-fold reductions in gauge speed, respectively.

GRIN was also more readily interpolated along the tested eight-bend deformation trajectory (Fig. 10c). With 33 calibration pins, the median held-out relative gauge error was $4.74 \times 10^{-5}$ for parabolic GRIN, compared with $3.69 \times 10^{-4}$ for near step index—a 7.8-fold reduction (each held-out gauge is aligned with its own transmission matrix, so this compares interpolation alone; branch recovery is treated in Section 2.6). The quartic-corrected profile gave an error of $3.84 \times 10^{-5}$, a 9.6-fold reduction relative to near step index. Together, the lower gauge speeds and interpolation errors support a smoother gauge trajectory for GRIN within this simulated comparison.

A local Nelder–Mead search around $\alpha = 2$ including a quartic term $\beta_4(r/a)^4$ found no consistent improvement on the parabola; near $\alpha = 2$ the combined score varies more with numerical detail than with the profile. Across $M = 1 \dots 8$ compounded bends, the graded-index core collapses the required pin count from exponential scaling in step index (8.0 → 29.6 → 1,499 pins at $M = 1,3,5$, exceeding the largest library tested, 16,384 pins, at $M = 8$) to low-degree polynomial scaling in parabolic GRIN (8.0 → 11.2 → 10.2 → 26.1), a reduction of more than 1,600× at $M = 8$. Part of that ratio comes from the bend sampler rather than the core: drawing both fibres' bends from one common sampler, step index needs 5.7 × graded-index's pins at $M = 8$ (215.7 against 38.1) rather than more than 600 ×. All counts are phase-aligned; Fig. 10d shows the common-sampler counts, and the ratios at M = 8 under the original sampler are lower bounds because step index is not covered within 16,384 pins. The direction of the result, a large and growing graded-index advantage, survives every check; its size depends on how bends are sampled.

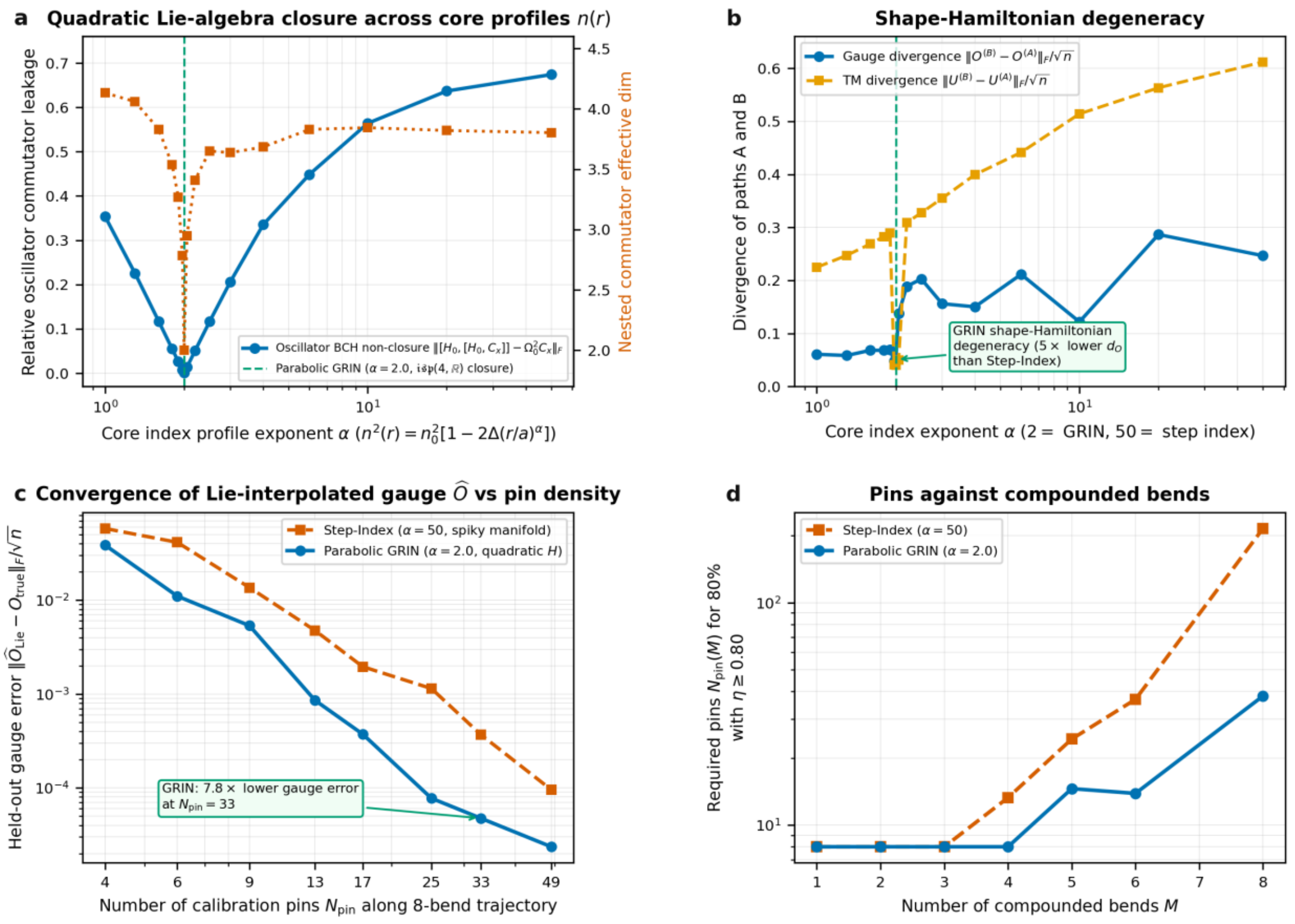

**Fig. 10.** Why a graded-index core keeps the manifold smooth. **a**, Sweeping the core profile $f(r/a) \propto (r/a)^\alpha$, the relative leakage of the bend Hamiltonian out of the quadratic oscillator algebra is minimised at the parabolic profile: 0.0013 at $\alpha = 2$ against 0.674 for a step index, a factor of about 540. **b**, The same parabolic core is far more degenerate to shape detail: pairs of eight-bend paths differing by 85% RMS curvature only in the part orthogonal to the pitch frequency move the gauge by only 0.048–0.052 and the transmission matrix by 0.041 for α = 1.97–2.00, against 0.247 and 0.612 at $\alpha = 50$, so a GRIN manifold responds to the bend's resonant curvature components rather than its exact shape. The transmission-matrix degeneracy is narrow in α (divergence below 0.05 only for α = 1.97–2.05, and at least 0.22 elsewhere), while the gauge divergence stays below 0.07 for every α ≤ 2. **c**, Lie interpolation therefore converges faster with pin density: at 9 pins the held-out gauge error is $5.3 \times 10^{-3}$ for parabolic GRIN, against $1.4 \times 10^{-2}$ for step index, and by 49 pins $2.4 \times 10^{-5}$ against $9.6 \times 10^{-5}$. **d**, The consequence for calibration cost when both fibres' bends are drawn from one common sampler: at M = 8, 80% coverage at η ≥ 0.80 needs 215.7 pins for step index against 38.1 for parabolic GRIN, 5.7× more. Phase-aligned counts from worksheet E15. Gauge errors in c are measured after aligning each held-out gauge with its own transmission matrix.

## 3. Discussion

This work establishes that single-ended reflection calibration of a non-unitary multimode fibre ($Q = U^T R U$) is governed by a precise mathematical quotient—a complex-orthogonal gauge $O \in O(n, \mathbb{C})$ modulo a discrete $n!\,2^n$ signed-permutation branch group $G = PS$—and uses simulations to turn that quotient geometry into engineering guidelines for flexible MMF endoscopy:

1. **Systematic Manifold Construction from Unknown Pins (**Fig. 1c**):** In practical deployment, a mechanical calibration scan records $(U_k, Q_k)$ across a sequence of bends without prior knowledge of which Takagi branch each numerical factorisation lands on. We showed that, on the one-parameter pyMMF family, proximal round-trip distances $d_Q(i, j)$ sort an unordered bend program (r = 0.999), and that matching each new pin's $(U_k, Q_k)$ against the nearest already-aligned bend's gauge $O_{k-1}$ recovers the exact signed permutation $G_{\mathrm{cross},k}$ onto a single aligned gauge manifold ($1.61 \times 10^{-15}$ sheet consistency), where the obtained pin gauges match Lie-interpolated gauges to $1.74 \times 10^{-4}$.
2. **Why Takagi Lie-Gauge Interpolation Beats Raw Database Averaging (**Fig. 2d, Fig. 7a**):** Direct Euclidean averaging of stored transmission matrices $\sum_m w_m U_{j_m}$ mixes rapidly oscillating phases and violates the measured quadratic constraint $U^T R U = Q_{\mathrm{new}}$. By contrast, Takagi Lie-gauge interpolation keeps the newly measured right Takagi factor $B_{Q,\mathrm{new}} = D_{Q,\mathrm{new}}^{1/2} V_{\mathrm{new}}^T$ exact and interpolates only the missing gauge $O$ in $\mathfrak{so}(n, \mathbb{C})$, reducing the held-out relative Frobenius error in $U$ by 65.5 × over nearest-pin look-up (median over the same one-parameter pyMMF family, with the global phase removed; at the 90th percentile the errors are 0.64 and 0.85). Interpolation across fibre shapes that vary in many parameters has been tested only in the simulated length models of Section 2.3. For a live state the branch must first be recovered from $Q$ (see below).
3. **Single-ended branch recovery (**Fig. 8, Table 1**):** A live measurement gives only $Q$, so its Takagi branch is unknown and cannot be read from its own transmission matrix, as it was for the held-out results above. The overlap phases between a live state and the nearest pin wind by several turns between pins (713° in the mode-solver model with 16 pins), so taking signs from the real part of the overlaps picks a harmful wrong branch for 30–68% of live states. Locating the live state from the reflector residual and its overlap phases, and judging each sign against the phase predicted from the unwrapped calibration, reduces this to 0.3% with 16–48 evenly spaced pins in the mode-solver model, and to at most 0.3% from 32 pins in the perturbative model. On the pyMMF bends it gives a median $F = 2.81 \times 10^{-4}$ leave-one-out ($1.74 \times 10^{-4}$ with the true branch), and on the nine pins of Fig. 2d it lowers the median from $2.28 \times 10^{-3}$ to $5.5 \times 10^{-4}$. A sign-confidence flag catches every harmful state in the perturbative model at 16 pins, but not unwrapping errors between widely spaced pins. When the fibre shape is measured, a sensor reading within 0.1 of the pin spacing does as well. The pins must be dense enough to follow each overlap phase between neighbours, and multi-segment fibres remain to be tested.

4. **Simulated Limits on Passive Database Methods (**Fig. 3, Fig. 4**):** Two fibre models agree that look-up cost grows exponentially with the number of independent bends and therefore fails beyond some length of free fibre, but place that length differently: at about 0.5 m in the perturbative model (about 2,472 pins for $\eta \geq 0.80$ at 0.5 m, and more than $10^5$ at 2 m) and between 2 and 5 m in the mode-solver step-index model. Its position depends on bend sensitivity and must be measured on real fibres. In the perturbative model, library size is governed by the number of independent mechanical degrees of freedom J (Fig. 3c,d): restricting a 2.0 m fibre to three hinge joints keeps it below 200 pins at $\eta \geq 0.80$.
5. **Full Dual-Polarisation Measurement & Distal Reflector Design (**Fig. 5a–d**):** In the vectorial simulator, elasto-optic torsional shear strain ($p_{44}$) scatters up to 73% of co-polarised light into the orthogonal polarisation under axial twist, so full $2N \times 2N$ polarimetric calibration is needed for accurate tracking ($\mathrm{cond}(U_{2N}) = 1.01, F = 0.029$ vs single-pol $U_{HH}$ cond $= 4.08, F = 0.697$). At the distal tip, a structured reflector such as a 4%-floored FZP or a patterned metasurface keeps $\eta \approx 0.90$ at $10^4$ photons per mode for single-wavelength pinned tracking (Fig. 5c). It works because it breaks the near-degenerate Takagi spectrum of a bare facet, which fails even without noise, not because it returns more light. Adding a three-wavelength Lee–Villiger dispersion sweep ($\lambda_0 \pm 0.10$ nm, Fig. 5d) breaks exact single-wavelength $O(n, \mathbb{C})$ gauge invariance to select the true gauge without prior pins.
6. **Waveguide Core Engineering via Quadratic Lie-Algebra Closure & Shape-Hamiltonian Degeneracy (**Fig. 10a–d**):** In the spectral core-profile solver, whereas step-index cores show exponential multi-bend pin growth due to non-commuting BCH leakage across $\mathfrak{u}(n)$ (0.674 leakage), parabolic GRIN ($\alpha = 2.00$) traps the bend Hamiltonian inside the quadratic oscillator Lie subalgebra $\mathfrak{isp}(4, \mathbb{R})$ (0.0013 leakage), rendering pitch-orthogonal 3D shape variations degenerate in $H_{\mathrm{eff}}$ ($5 \times$ lower gauge divergence), reducing Lie-interpolated gauge error $7.8 \times$, and reducing multi-bend pin counts (5.7× fewer than step index at eight bends under a common bend sampler, and more under the original sampler).
7. **What the Measured Data Show (**Fig. 6**):** On 139 measured transmission matrices of a 5 m step-index fibre, a Tikhonov-regularised reflector score identifies the nearest remaining calibration for 84% of held-out targets, against 3% with the exact inverse, so reflector de-embedding can select calibrations on real, poorly conditioned matrices. The recorded bends are too far apart for any method, including ours, to reconstruct a held-out bend: the nearest measured neighbour of each transmission matrix differs from it by a median F = 0.94 ($\eta \approx 0.31$), and even the closest pair by F = 0.67 (η ≈ 0.61); a finely sampled experiment is therefore the decisive next test.

Limitations. Both length models are scalar and bend-only; they omit polarisation coupling, twist, core ellipticity, microbending and random index variations, any of which may change bend sensitivity in either direction. The perturbative model is phenomenological: its couplings are random matrices scaled to pyMMF, its straight-segment phases are wrapped to 2π before being combined with bend terms that do not commute with them, its second-order bend terms are comparable to the first-order ones, and its mode-dependent loss, which creates the gauge, is an assumed 0.008 $(g/g_max)^2$ per metre. The mode-solver model computes its couplings from mode overlaps, but on 45 recorded shapes of the measured 5 m fibre it decorrelates faster than the real fibre and does not reproduce which pairs are similar (rank correlation 0.06), and its libraries stop at 1,024 pins. Neither model is calibrated against measured decorrelation, so every length quoted applies to the model named with it, and the range between the two models indicates rather than bounds the behaviour of a real fibre. The coverage figures are for sampled held-out fibres, not guarantees over every shape. Degeneracy handling must be judged against the noise resolution of the Takagi values rather than a fixed tolerance. Single-ended branch recovery needs pins dense enough to follow the overlap phases of the Takagi vectors; where they are not (the tightest pyMMF bends, the perturbative model below about 32 pins) it fails, and the sign-confidence flag catches most but not all of these failures. Patterned facet reflectors also mix modes. With the off-centre zone plate projected on the real mode fields of the mode-solver model, single-ended branch recovery did as well as with the idealised reflector, but the interpolation error with the true branch was about 18 times larger at 16 pins (median $F = 1.2 \times 10^{-2}$ against $6.8 \times 10^{-4}$), so the accuracies reported here may not carry over unchanged to a fabricated reflector.

---

## Appendix A: Methods

### A.1 Batched GPU Autonne–Takagi Factorisation and Degeneracy-Safe Subspace Reduction

For a complex-symmetric matrix $A = A^T \in \mathbb{C}^{n\times n}$, `takagi_factor` (`pinning_manifold/takagi_gpu.py`) forms the Hermitian positive-semidefinite product $B = \frac{1}{2}\left(A\overline{A} + \left(A\overline{A}\right)^H\right)$, computes its real eigenvalues $\mu_k \geq 0$ and unitary eigenvectors $V_0 \in U(n)$ in `complex128` via `torch.linalg.eigh`, and sorts singular values $s_k = \sqrt{\mu_k}$ in descending order. Each non-degenerate eigenvector $v_k$ is phase-aligned via $V_{:,k} = v_k \exp\left(\frac{i}{2}\arg(v_k^H A \overline{v_k}/s_k)\right)$. Within any degenerate or near-repeated singular-value cluster $\mathcal{J}$ ($|s_i - s_j|/s_i \leq 10^{-9}$), the unitary symmetric subspace matrix $M_\mathcal{J} = s_\mathcal{J}^{-1} V_\mathcal{J}^H A \overline{V_\mathcal{J}}$ is diagonalised via the real orthogonal eigenvectors $Q_\mathbb{R} \in O(|\mathcal{J}|, \mathbb{R})$ of the commuting real symmetric combination $Re\left(e^{-i\theta} M_\mathcal{J}\right) + \eta Im\left(e^{-i\theta} M_\mathcal{J}\right)$, guaranteeing machine-precision Takagi residuals across both simple and repeated spectra.

### A.2 Whitened Hungarian Branch Alignment, Cross-Bend Program Sorting and $\mathfrak{so}(n, \mathbb{C})$ Interpolation

Given $X_0 \approx Y_0 G^T$ with $G = PS \in P_n \ltimes \{\pm 1\}^n$, `recover_signed_permutation_g` normalises the columns of $X_0$ and $Y_0$, computes $C = \hat{X}^H \hat{Y}$, solves the Hungarian linear assignment problem $\min_{\pi \in S_n} \sum_{k=1}^{n} \left(1 - \left|C_{k,\pi(k)}\right|\right)$, and sets $G_{k,\pi(k)} = sign\left(Re\, C_{k,\pi(k)}\right)$; align_gauge_to_reference(O, O_ref) calls it with the reference first, $X = O_{\mathrm{ref}}$, $Y = O$, and returns the correction $G^T$, so that $OG^T \approx O_{\mathrm{ref}}$. Given an unordered mechanical calibration program of pins $\{(U_k, Q_k)\}_{k=1}^{N_{pin}}$ with arbitrary Takagi branches $\left(V_{raw,k}, D_{Q,raw,k}\right)$, the pins are ordered by greedy nearest-neighbour traversal on $d_Q(i,j) = \| Q_i - Q_j \|_F / \| Q_i \|_F$ starting from the diameter endpoint $\mathrm{argmax}_{i,j} d_Q(i,j)$. At each step $k$, the raw gauge $O_{raw,k} = D_R^{1/2} W^T U_k \left(D_{Q,raw,k}^{1/2} V_{raw,k}^T\right)^{-1}$ is aligned to the nearest already-aligned bend's gauge $O_{k-1}$ via $G_{cross,k} = align_gauge_to_reference\left(O_{raw,k}, O_{k-1}\right)$, placing $O_k = O_{raw,k} G_{cross,k}$ onto the gauge manifold. Relative gauges $M_k = O_{ref}^{-1} O_k \in SO(n, \mathbb{C})$ are mapped to $\mathfrak{so}(n, \mathbb{C})$ via $L_k = \frac{1}{2}(\log M_k - (\log M_k)^T)$.

**Live states.** For a live state only $Q_{live}$ is measured, so $V_{live}$, $D_{Q,live}$ and $B_{Q,live}$ are known but $U_{live}$ is not, and the whitened matrix $Y = UB_Q^{-1}$ of the alignment above cannot be formed. The branch must instead be inferred by comparing the live Takagi vectors with those of the pins, through the overlaps $C = V_{pin}^H V_{live}$. The assignment is the same Hungarian problem on $|C|$, but the signs cannot simply be taken from Re $C$, because the overlap phases turn by more than 90° between pins. The single-ended pipeline (single_ended.py; Section 2.6) therefore first locates the live state between the pins from the reflector residual and its overlap phases, predicts each overlap phase there from the unwrapped calibration phases, and takes each sign against that prediction. The gauge is then interpolated at the estimated position as above, and $U_{live} = A_R^{-1} O B_{Q,live}$.

### A.3 Standardized 6-Method Single-Ended Reconstruction Benchmark

For a pinned calibration library $\{U_j\}_{j=1}^{N_{pin}}$ ($\| U_j \|_F = \sqrt{n}$) and a held-out state with measured $Q_{test} = U_{test}^T R U_{test}$, every library candidate is scored without ground-truth leakage by solving $U_j^T R_{hat,j} U_j = Q_{test}$ and evaluating $\epsilon_R(j) = \| R_{hat,j} - R \|_F / \| R \|_F$. The top $k = 4$ neighbours $j_1, \ldots, j_k$ with soft weights $w_m \propto exp\left(-\left(\epsilon_R(j_m) - \epsilon_R(j_1)\right)/\tau\right)$ are passed identically to `rhat_nearest`, `raw_tm_euclidean`, `raw_tm_generator`, `takagi_lie` and `physics_shape_proposal`, while `oracle_nearest` reports $\min_j \| U_j - U_{test} \|_F / \| U_{test} \|_F$.

### A.4 Multi-Length WLC, Mechanical Multi-Join, Vectorial Twist/Strain and Lee–Villiger Simulators

In `fibre_simulator_gpu.py`, `PerturbativeBentMMFSimulatorGPU` uses the equally spaced mode-group propagation constants of a parabolic core, with random dipole bend operators $C_x, C_y$ and quadratic operators $Q_{xx}, Q_{yy}, Q_{xy}$ across worm-like chain curvature realisations ($L \in [0.05, 5.00]$ m) and epoxied multi-join constraints ($J \in \{1,2,3,5,10\}$ joints on $L = 2.00$ m). `VectorialPolarimetricTwistThermalSimulatorGPU, built on the`

`perturbative model`, models the $2N \times 2N$ spin-orbit Hamiltonian combining Ulrich–Simon photoelastic shear strain ($g_{elasto} = 0.15$), core ellipticity ($e_{core} = 1.5 \times 10^{-3}$) and thermo-optic drift $\Delta T$. `LeeVilligerMultiWavelengthSimulatorGPU` implements symmetric distributed/distal second-order chromatic dispersion $U(\Delta\lambda) = P_{1/2}(\Delta\lambda)U_0P_{1/2}(\Delta\lambda)$ using the physical mode propagation constants $\boldsymbol{\beta}_0$ from `M0.mat`.

The perturbative model is phenomenological. C_x and C_y are random complex matrices whose entry sizes follow the |Δg| = |Δℓ| = 1 selection rule, with weaker |Δg| ≤ 2 and residual terms, normalised so that $\kappa = 1\ m^{-1}$ over 0.10 m reproduces the mixing rate of the pyMMF bends; Q_αβ = 0.08 C_αC_β; straight-segment phases βΔz are taken modulo 2π; and mode-dependent loss is $\gamma_g = 0.008\ (g/g_max)^2\ m^{-1}$. Because these couplings are neither derived from a parabolic index profile nor restricted to |Δg| = 1, the model does not have the bend insensitivity of a GRIN fibre; it is a phenomenological fibre with a GRIN mode spectrum. The mode-solver model (physical_sim_numpy.PhysicalBentFibre) solves the scalar guided modes of a step-index core (radius 25 µm, NA 0.22, n = 1.457; about 120 simulated and 55 observed modes) on a transverse grid, builds each bent segment from the dipole and quadrupole overlaps of those modes with the conformal-mapping bend perturbation, and propagates worm-like-chain bend programmes identical to those of the perturbative model; Fig. 4 is drawn by make_fig_two_models.py, with 50 observed modes in panel a to match pyMMF and 55 in panels b and c, from its stored results.

**Reflector.** Unless stated otherwise, the simulations use an idealised reflector, diagonal in the modal basis, $R = \mathrm{diag}(r_k e^{i\varphi_k})$ with $r_k$ uniform on [0.35, 1] and random phases $\varphi_k$ (make_diagonal_reflector): it does not mix modes, and its Takagi values are distinct, as the design rule of Section 2.4 requires. The patterned facet reflectors of Sections 2.4 and 2.5 are instead projected on the fibre's real mode fields.

### A.5 Experimental 5 m step-index MMF Intake and SVD Modal Bridge

In `experimental_bridge.py`, 2D spline coordinates (`spline_path.mat`, stored y before x; 5.758 px $cm^{-1}$ from `calibration.mat`), rectangular transmission matrices `TM.mat` ($A_k \in \mathbb{C}^{32,017\times145}$), intensity reflection matrices `RM.mat` and pupil masks `pupils.mat` (`203 x 203` mask `bl` with 32,017 active pixels) are loaded from `Experiment2-20260330` of the public dataset[29]. The canonical distal modal basis $W_{dist} \in \mathbb{C}^{32,017\times145}$ is extracted via GPU SVD, mapping $A_k \mapsto U_k = W_{dist}^H A_k \in \mathbb{C}^{145\times145}$. In the released result files the stored keys `*_strehl` and `good_fraction_s08/s04` hold $\eta = max(0,1 - F^2/2)^2$ and coverage at $\eta \geq 0.80/0.40$ respectively. Every comparison between matrices in this paper uses the phase-aligned $F$, computed as $F^2 = 2 - 2\left|tr(\widehat{U}^H U)\right|/n$ for $\| \widehat{U} \|_F = \| U \|_F = \sqrt{n}$; the regulariser added before factorising measured round trips is applied in phase, $U + 0.02\, e^{i\arg trU} I$, because at a fixed phase it changed the Takagi values by up to 50% under a change of global phase.

### A.6 Gauss–Laguerre Spectral Radial Helmholtz Solver and Core Profile $\mathbf{n(r)}$ Optimization

In `experiments/e7_grin_vs_step_index_and_optimal_core.py`, `RadialHelmholtzMMFCoreBuilder` solves the radial Helmholtz eigenproblem across core profiles $f(r/a) = (1 - \beta_4)(r/a)^\alpha + \beta_4(r/a)^4$ for each azimuthal order $\ell$ by expanding radial modes in an exact orthonormal generalised Laguerre basis $\Phi_{p,|\ell|}(x)$ integrated via 96-point Gauss–Laguerre quadrature ($x = V(r/a)^2$), avoiding finite-difference centrifugal singularities at $r = 0$. From the top $n = 36$ guided modes, exact dipole ($C_x, C_y$) and quadrupole ($Q_{xx}, Q_{yy}, Q_{xy}$) coupling matrices and non-paraxial propagation constants $\beta_{p,\ell} = \sqrt{k_0^2 n_{core}^2 - k_{\perp,p,\ell}^2}$ are assembled on GPU to evaluate oscillator double-commutator closure $\| [H_0, [H_0, C_x]] - \Omega_0^2 C_x \|_F$, pitch-orthogonal shape-Hamiltonian degeneracy and Lie-interpolation convergence vs $N_{pin}$. The local search around α = 2 reported in Section 2.7 is in numpy_notebooks/check_quartic_neighbourhood.py.

### A.7 Use of Artificial Intelligence

Large language models (Anthropic Claude Opus 5.5, OpenAI GPT-6.1 Sol, and Google Gemini Argon 2) were used throughout this work as research and writing assistants: to write and debug simulation and analysis code, to propose and implement computational checks, to run analyses and to draft and edit text. No figures were generated by image-generation models; all figures are plots produced by the released code. The authors conceived the approach, directed all analyses, checked the code, derivations, and numerical results, and take full responsibility for the content of the manuscript. All code, including AI-assisted code, is available as described under Data and Code Availability. No language model is listed as author.

## Appendix B: Exact Autonne–Takagi quotient geometry of single-ended reflection

### B.1 Algebraic Statement of the Single-Ended Inverse Problem

Consider a reciprocal linear optical channel with one-way forward transmission matrix $U \in GL(n, \mathbb{C})$ mapping $n$ proximal input modes to $n$ distal output modes. By optical Lorentz reciprocity, the backward propagation from the distal facet to the proximal facet in the transpose-dual modal basis is $U^T$. When a linear reciprocal distal reflector with complex-symmetric reflection matrix $R = R^T \in GL(n, \mathbb{C})$ terminates the distal facet, the proximal round-trip reflection matrix is

$$Q = U^T R U, \qquad Q^T = Q \in \mathbb{C}^{n \times n}.$$

Because $Q$ is complex symmetric, its complex degrees of freedom number $n(n+1)/2$, whereas the general non-unitary transmission matrix $U \in GL(n, \mathbb{C})$ possesses $n^2$ complex degrees of freedom. Consequently, single-ended measurement of $Q$ at a single wavelength with a known reflector $R$ leaves $n^2 - n(n+1)/2 = n(n-1)/2$ complex degrees of freedom ($n(n-1)$ real degrees of freedom) unresolved.

### B.2 Autonne–Takagi Factorisation and the Complex-Orthogonal Gauge Group $\mathrm{O}(\mathrm{n}, \mathbb{C})$

By the Autonne–Takagi theorem, any complex-symmetric matrix $A = A^T \in \mathbb{C}^{n \times n}$ admits a factorisation

$$A = U_A \Sigma_A U_A^T,$$

where $U_A \in U(n)$ is unitary and $\Sigma_A = diag(\sigma_1, \dots, \sigma_n)$ with $\sigma_1 \geq \sigma_2 \geq \cdots \geq \sigma_n \geq 0$ contains the Takagi singular values (the non-negative square roots of the eigenvalues of the Hermitian positive-semidefinite matrix $A\overline{A}$). Applying Autonne–Takagi factorisation separately to the calibrated distal reflector $R$ and the measured proximal round-trip matrix $Q$ yields

$$R = W D_R W^T, \qquad Q = V D_Q V^T,$$

where $W, V \in U(n)$ and $D_R, D_Q$ are positive diagonal matrices when $R$ and $U$ are invertible. Defining the symmetric square-root factors

$$A_R = D_R^{1/2} W^T \quad (\text{so } R = A_R^T A_R), \qquad B_Q = D_Q^{1/2} V^T \quad \left(\text{so } Q = B_Q^T B_Q\right),$$

the round-trip relation $Q = U^T R U$ becomes

$$B_Q^T B_Q = (A_R U)^T (A_R U) \Leftrightarrow \left(A_R U B_Q^{-1}\right)^T \left(A_R U B_Q^{-1}\right) = I_n.$$

Therefore, the matrix

$$O \equiv D_R^{1/2} W^T U \left(D_Q^{1/2} V^T\right)^{-1} \in O(n, \mathbb{C}) = \{M \in \mathbb{C}^{n \times n} : M^T M = I_n\}$$

belongs to the complex orthogonal group $O(n, \mathbb{C})$, and every transmission matrix $U$ consistent with $(R, Q)$ is parameterised exactly by

$$U = (W^T)^{-1} D_R^{-1/2} O D_Q^{1/2} V^T = \overline{W} D_R^{-1/2} O D_Q^{1/2} V^T.$$

### B.3 Degeneracy-Safe Subspace Takagi Reduction

When two or more Takagi singular values of $A = A^T$ are equal or nearly equal ($\sigma_i \approx \sigma_j$, as occurs for an unpatterned bare Fresnel facet where $D_R \approx r_0 I_n$), standard eigendecomposition of $B = A\overline{A}$ returns an arbitrary unitary basis $V_\mathcal{J} \in \mathbb{C}^{n \times m}$ for the $m$-dimensional degenerate eigenspace $\mathcal{J}$ with eigenvalue $\sigma^2$. On this subspace, the matrix

$$M_\mathcal{J} = \frac{1}{\sigma} V_\mathcal{J}^H A \overline{V_\mathcal{J}}$$

is simultaneously unitary ($M_\mathcal{J} M_\mathcal{J}^H = I_m$) and complex symmetric ($M_\mathcal{J}^T = M_\mathcal{J}$). Consequently, the real and imaginary parts of $e^{-i\theta} M_\mathcal{J}$ are commuting real symmetric matrices. Diagonalising a generic real linear combination $H_{\text{real}} = Re\left(e^{-i\theta} M_\mathcal{J}\right) + \eta Im\left(e^{-i\theta} M_\mathcal{J}\right)$ via a real orthogonal matrix $Q_{\text{real}} \in O(m, \mathbb{R})$ and adjusting the $m$ column phases by $\phi_k = \frac{1}{2} \arg(Q_{\text{real}}^T M_\mathcal{J} Q_{\text{real}})_{kk}$ yields $U_\mathcal{J} = Q_{\text{real}} diag\left(e^{i\phi_k}\right)$ satisfying $M_\mathcal{J} = U_\mathcal{J} U_\mathcal{J}^T$ to machine precision ($< 10^{-15}$).

### B.4 Discrete Signed-Permutation Branch Ambiguity $\mathbf{G = PS}$

When the Takagi singular values of $Q$ are distinct and positive, the ordered Takagi factor $V$ is unique up to right-multiplication by a diagonal sign matrix $S = diag(s_1, \dots, s_n)$ with $s_k \in \{+1, -1\}$. If singular values cross or are re-indexed along a deformation trajectory by a permutation matrix $P \in S_n$, the general branch ambiguity of the Takagi pair $(V, D_Q)$ is generated by the finite signed-permutation group

$$G = \{PS: P \in S_n,\ S = diag(\pm 1, \dots, \pm 1)\} \le O(n, \mathbb{R}) \le O(n, \mathbb{C}), \qquad |G| = n!\, 2^n.$$

Under a branch replacement $(V, D_Q) \mapsto (\tilde{V}, \tilde{D}_Q) = (VG, G^T D_Q G)$, invariance of the physical transmission matrix $U$ requires the gauge matrix to transform covariantly as

$$\tilde{O} = OG \Rightarrow O = \tilde{O}G^T.$$

Thus, the space of unaligned Takagi gauges observed over a physical fibre deformation domain $\mathcal{M}_{\text{phys}}$ consists of $n!\, 2^n$ disjoint isometric copies $\mathcal{M}_O \cdot G$ of the single continuous **aligned gauge manifold** $\mathcal{M}_O \subset O(n, \mathbb{C})$.

## Appendix C: Whitened Hungarian branch recovery and Lie-algebra interpolation

### C.1 Exact Branch Recovery from a Reference Pin

Suppose $U_{\text{ref}}$ is a known calibration pin near $U$ on the deformation manifold, and let $B_Q = D_Q^{1/2} V^T$ be a Takagi square-root factor of $Q = U^T R U$ in an arbitrary branch $\tilde{V} = VG$. Forming the whitened matrices

$$X = U_{\text{ref}} B_{Q,\text{ref}}^{-1} \approx \overline{W} D_R^{-1/2} O, \qquad Y = U_{\text{wrong}} B_Q^{-1} = \overline{W} D_R^{-1/2} OG,$$

we satisfy $X \approx YG^T$. Normalising the columns of $X$ and $Y$ to unit Euclidean norm ($\hat{X}_{:,k} = X_{:,k} / \| X_{:,k} \|_2$, $\hat{Y}_{:,j} = Y_{:,j} / \| Y_{:,j} \|_2$) and forming the cross-correlation matrix $C = \hat{X}^H \hat{Y}$, the optimal permutation $\hat{\pi} \in S_n$ is obtained in $O(n^3)$ time via the Hungarian linear sum assignment algorithm:

$$\hat{\pi} = arg \min_{\pi \in S_n} \sum_{k=1}^{n} \left(1 - |C_{k,\pi(k)}|\right), \qquad \hat{s}_k = sign\left(Re\left(C_{k,\hat{\pi}(k)}\right)\right) \in \{+1, -1\}.$$

Setting $\hat{G}_{k,\hat{\pi}(k)} = \hat{s}_k$ yields the branch-corrected gauge $O_{\text{corr}} = (OG)\hat{G}^T = O$. On all 39 well-conditioned pyMMF bends (`2.mat`–`40.mat`, $n = 50$), this procedure recovers $\hat{G} = G_{\text{true}}$ with zero errors and median relative reconstruction residual $F = 1.339 \times 10^{-15}$ (compared to $F = 1.443$ before branch alignment). This recovery requires the state's own transmission matrix to form $Y$, so it applies to the calibration pins and to the oracle comparisons; a live state, for which only $Q$ is measured, is handled differently (Appendix A.2, Section 2.6).

### C.2 Projection and Geodesic Interpolation in $\mathfrak{so}(\mathbf{n}, \mathbf{C})$

The Lie algebra of the complex orthogonal group $O(n, \mathbb{C})$ is the space of complex skew-symmetric matrices:

$$\mathfrak{so}(n, \mathbb{C}) = \{L \in \mathbb{C}^{n \times n}: L^T = -L\}, \qquad \dim_{\mathbb{C}} \mathfrak{so}(n, \mathbb{C}) = \frac{n(n-1)}{2}.$$

Given branch-aligned pinned gauges $O_1, \dots, O_k \in SO(n, \mathbb{C})$ around a local reference pin $O_1$, each relative gauge $M_m = O_1^{-1} O_m \in SO(n, \mathbb{C})$ is mapped to $\mathfrak{so}(n, \mathbb{C})$ via the principal matrix logarithm followed by skew-symmetric projection $\Pi_{\mathfrak{o}}(X) = \frac{1}{2}(X - X^T)$:

$$L_m = \Pi_{\mathfrak{o}}\left(\log(O_1^{-1} O_m)\right) \in \mathfrak{so}(n, \mathbb{C}).$$

Across the pyMMF bend sequence, the raw principal logarithm already satisfies $\| L_{\text{raw}} + L_{\text{raw}}^T \|_F / \| L_{\text{raw}} \|_F = 1.05 \times 10^{-13}$, and exponentiating the projected skew-symmetric matrix reproduces $O_m$ to relative error $1.67 \times 10^{-14}$. Weighted geodesic interpolation with weights $w_m \geq 0$ ($\sum_{m=1}^{k} w_m = 1$) takes the form

$$\hat{O} = O_1 \exp\left(\sum_{m=1}^{k} w_m \, \Pi_{\mathfrak{o}}\left(\log(O_1^{-1} O_m)\right)\right) \in SO(n, \mathbb{C}),$$

which guarantees $\hat{O}^T \hat{O} = I_n$ and $\widehat{U}^T R \widehat{U} = Q_{\text{new}}$ identically.

## Appendix D: Perturbative single-bend Hamiltonian and asymptotic composition

### D.1 Coupled-Mode Curvature Expansion

For a weakly guiding graded-index or step-index MMF segment of length $\Delta z$ bent with curvature $\kappa = 1/R_{\text{bend}}$ in azimuth plane $\phi$, conformal mapping of the curved waveguide coordinates transforms the scalar Helmholtz equation into an effective coupled-mode Hamiltonian:

$$H(\kappa, \phi) = diag(\boldsymbol{\beta}_0 + i\boldsymbol{\gamma}_0) + \kappa\left(\cos\phi \, C_x + sin\phi \, C_y\right) + \kappa^2\left(\cos^2\phi \, Q_{xx} + \sin^2\phi \, Q_{yy} + sin(2\phi) \, Q_{xy}\right) + \mathcal{O}(\kappa^3),$$

where $(C_x)_{mn} = k_0 n_{\text{core}}(1 - p_e)\langle\psi_m|x|\psi_n\rangle$ and $\left(C_y\right)_{mn} = k_0 n_{\text{core}}(1 - p_e)\langle\psi_m|y|\psi_n\rangle$ obey the dipole selection rule $|\Delta g| = 1, |\Delta \ell| = 1$ between adjacent principal mode groups, and $Q_{xx}, Q_{yy}, Q_{xy}$ capture second-order geometric/elasto-optic wavefront compression ($|\Delta g| \in \{0,2\}$).

### D.2 Multiplicative Composition of $(\mathbf{O}, \mathbf{B})$ and Baker–Campbell–Hausdorff Spikiness

When a fibre of length $L$ is composed of $M = L/\ell_c$ independent bent segments,

$$U^{(M)}(\boldsymbol{\kappa}) = \prod_{m=1}^{M} \exp\left(i\ell_c H(\kappa_m, \phi_m)\right),$$

the effective generator $\log U^{(M)}$ accumulates nested Baker–Campbell–Hausdorff (BCH) commutators:

$$\log(e^{X_M} \cdots e^{X_1}) = \sum_{m=1}^{M} X_m + \frac{1}{2}\sum_{m>j} \left[X_m, X_j\right] + \frac{1}{12}\sum_{m,j,k} \left[X_m, \left[X_j, X_k\right]\right] + \cdots$$

Because $[C_x, C_y] \neq 0$ and $[diag(\boldsymbol{\beta}_0), C_{x,y}] \neq 0$, in the perturbative model composing multiple bends increases the Lie-algebra velocity magnitude $v(s) = \| \left(O^{(M)}\right)^{-1} \partial_s O^{(M)} \|_F$ from 0.021 at $M = 1$ ($L = 0.10$ m) to 20.84 at $M = 12$ ($L = 1.20$ m) and generates localized high-curvature ("spiky") transitions where the Jacobian $\| \partial O^{(M)} / \partial \boldsymbol{\kappa} \|$ peaks.

## Appendix E: Metric entropy and the scaling of pinning density

Let $\mathcal{M}(L) \subset GL(n, \mathbb{C})$ denote the transmission-matrix manifold of a multimode fibre of physical length $L$ deformed under a random curvature process $\kappa(s)$ with persistence length $\ell_p$ and effective intrinsic dimension $d_{\text{eff}}(L) \propto d_0(L/\ell_p)$ for a free elastic spline (or $d_{\text{eff}} = J \cdot d_{\text{joint}}$ for a mechanically constrained fibre with $J$ discrete joints).

To guarantee that a held-out state $U \in \mathcal{M}(L)$ has a pinned library state $U_j$ within normalized Frobenius error $F \leq \epsilon$ (where $\epsilon = \sqrt{2\left(1 - \sqrt{S_{\text{target}}}\right)}$ equals 0.4595 for $S_{\text{target}} = 0.80$ and 0.8574 for $S_{\text{target}} = 0.40$), Kolmogorov metric entropy requires a covering number scaling as

$$N_{\text{pin}}(L, \epsilon) \asymp \left(\frac{R_{\text{eff}}(L)}{\epsilon_{\text{eff}}}\right)^{d_{\text{eff}}(L)} \Rightarrow \log_{10} N_{\text{pin}}(L, \epsilon) \approx a(\epsilon) + \frac{L}{L_0(\epsilon)}.$$

When Takagi Lie-gauge interpolation is used instead of zeroth-order nearest-pin lookup, local third-order spline / geodesic interpolation reduces the error on a grid of spacing $h$ from $\mathcal{O}(h)$ to $\mathcal{O}(h^p)$ ($p \geq 2$), enlarging the effective tolerance radius $\epsilon_{\text{eff}}$ and reducing the required pin count by factors of $10\times$–$100\times$ in the resolved regime of the perturbative model ($L \leq 1.0$ m or $J \leq 5$ constrained joints).

## Appendix F: Vectorial 2N × 2N elasto-optic torsional twist and single-polarisation truncation

In a circularly symmetric ideal waveguide without mechanical strain, an axial twist $\tau = d\theta_{\text{twist}}/dz$ acts trivially on linearly polarised modes. In a real silica multimode fibre, two physical mechanisms couple twist strongly to the transmission matrix: 1. Ulrich–Simon Photoelastic Shear Strain ($p_{44}$): Twisting a cylindrical silica rod at rate $\tau$ produces shear strains $\gamma_{xz} = -\tau y$ and $\gamma_{yz} = \tau x$, which perturb the off-diagonal permittivity tensor via the elasto-optic coefficient $p_{44} = (p_{11} - p_{12})/2 \approx -0.075$. This induces circular birefringence with elasto-optic coupling constant $g_{\text{elasto}} = -n_0^2 p_{44} \approx 0.15$ and spin-orbit coupling $\tau g_{\text{elasto}}(\sigma_z \otimes L_z)$. 2. Intrinsic Core Ellipticity and Bend Stress Birefringence: Core ellipticity $e_{\text{core}} \sim 1.5 \times 10^{-3}$ and bend-induced lateral stress create a spatially varying linear birefringence operator $\sigma_z \otimes E_{\text{quad}} + \sigma_x \otimes C_{\text{bend}}$ whose principal axes rotate with the twisted fibre frame.

When the full $2N \times 2N$ vectorial transmission matrix $U_{2N}$ is measured, $U_{2N}$ remains nearly unitary ($cond(U_{2N}) = 1.01$) and its Takagi gauge $O_{2N} \in O(2N, \mathbb{C})$ evolves smoothly along $\tau$, yielding $F = 0.029$ ($\eta = 0.9991$) under 9-pin Takagi Lie-gauge interpolation across $\tau \in [-2.5, +2.5]$ rad/m. Conversely, when only a single input and single output linear polarisation block $U_{HH} = P_H U_{2N} P_H^T \in \mathbb{C}^{N\times N}$ is recorded, an average of 46.4% (and up to 73.2%) of the optical energy is lost into the unmeasured orthogonal polarisation $U_{VH}$, inflating $cond(U_{HH})$ to 4.08, inducing Takagi singular-value crossings in $Q_{HH} = U_{HH}^T R U_{HH}$, and degrading single-polarisation interpolation to $F = 0.697$ ($\eta = 0.573$).

## Appendix G: Supplementary figures

### G.1 Figure S1

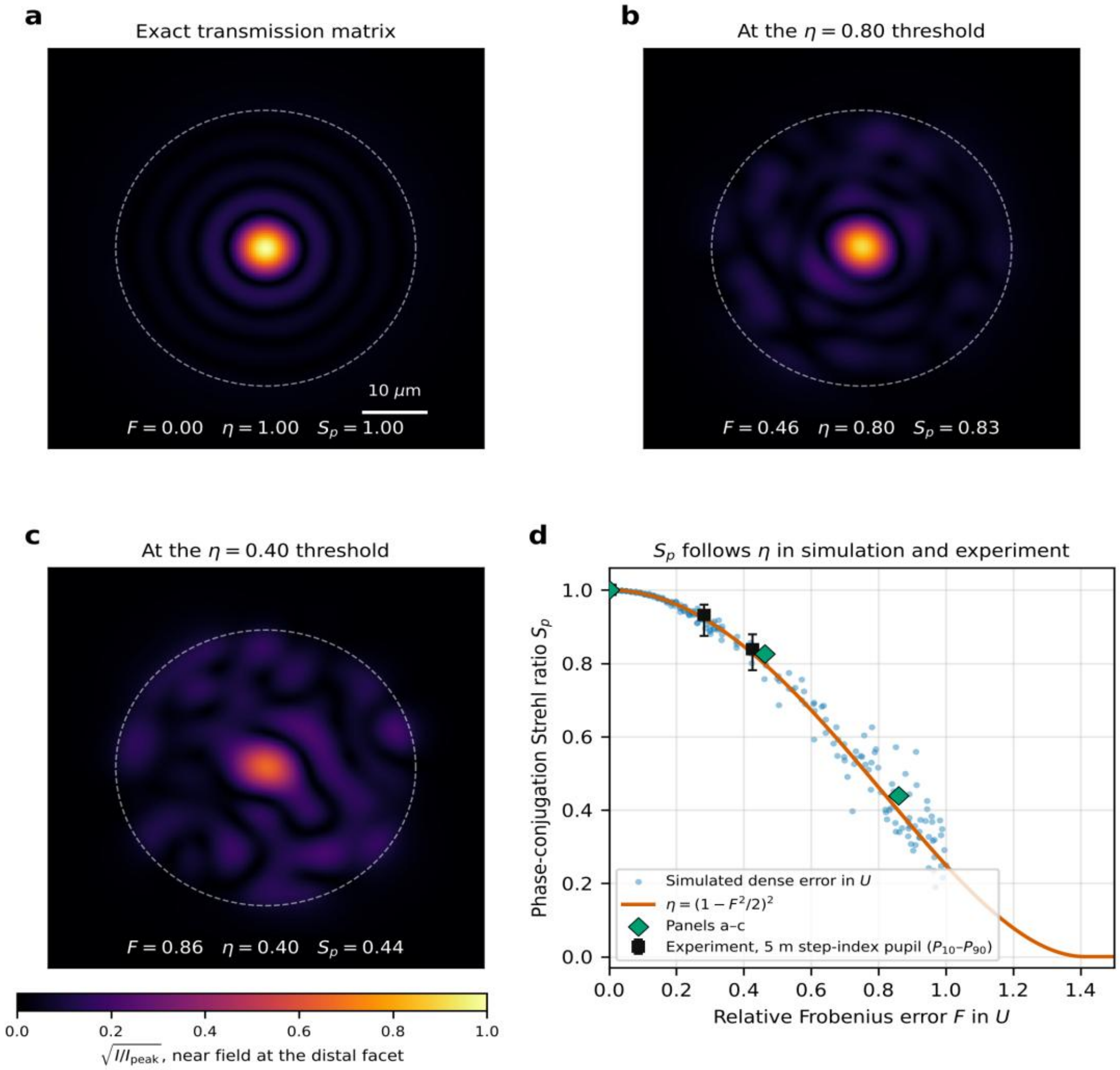

**Fig. S1.** What a given $F$ costs optically. Light is focused to the centre of the distal facet by phase conjugation against an estimate, $\mathbf{e}_{in} \propto \widehat{U}^{\dagger}\mathbf{t}$, and propagation through the truth, $\mathbf{e}_{out} = U\mathbf{e}_{in}$; the phase-conjugation Strehl ratio $S_p$ is the delivered peak intensity divided by the ideal $\widehat{U} = U$ value. The near field is rendered in a step-index LP basis fitted to the propagation constants of bend 2 ($V = 14.38$, NA $= 0.060$, core radius 24.2 $\mu$m, $\lambda =$ 632.8 nm), reproducing all 50 values of $n_{eff}$ to an rms of $6.9 \times 10^{-6}$. Intensities use a $\sqrt{\cdot}$ stretch; the dashed circle marks the core boundary. **a**, Exact estimate — the tightest spot the 50 guided modes can form. **b**, At the $\eta = 0.80$ threshold used throughout the main text, $F = 0.46$ gives $S_p = 0.83$: a tight core with a visible speckle haze. **c**, At $\eta = 0.40$, $F = 0.86$ gives $S_p = 0.44$; the focus survives but the background approaches the side lobes. **d**, Across 208 draws of a dense Gaussian error in $U$, $S_p$ follows $\eta = (1 - F^2/2)^2$ with no fitting, as do the four pupil values from the measured 5 m step-index fibre ($P_{10}$–$P_{90}$ bars). An error concentrated in a few modes at the same $F$ need not follow this curve.

## G.2 Figure S2

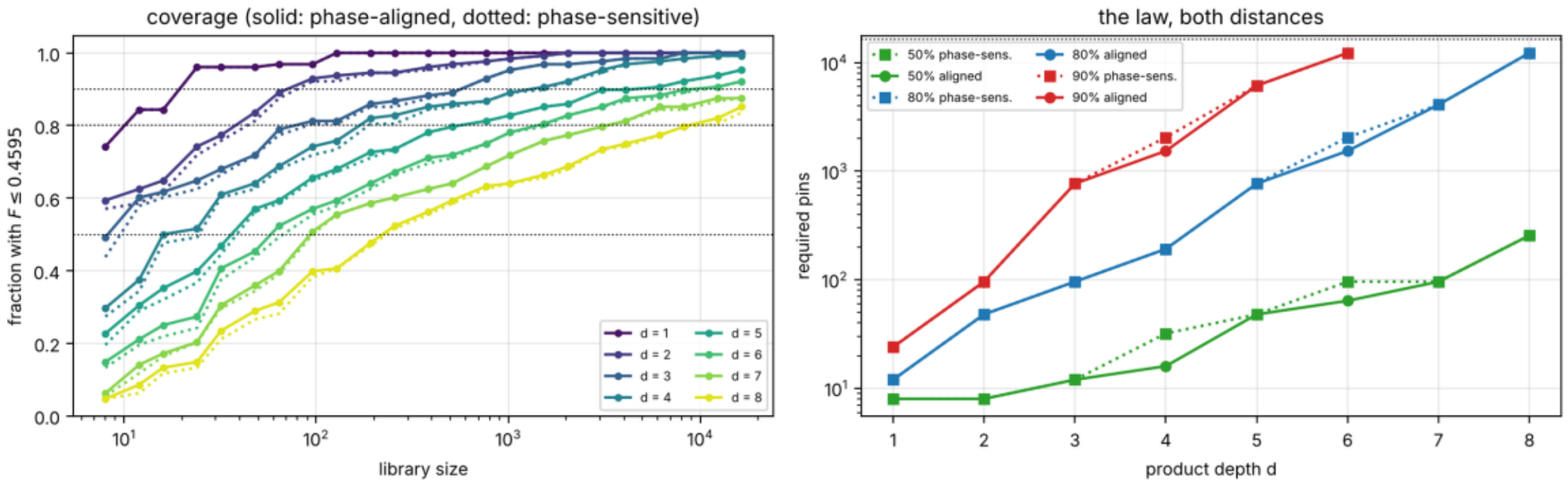

**Fig. S2.** Robustness of the product-depth scaling law to the global phase. Left, coverage against library size for products of d = 1–8 randomly drawn pyMMF bend matrices (solid, phase-aligned; dotted, phase-sensitive). Right, the library size needed for 50%, 80% and 90% coverage against d. The multiplier per bend is 2.61 phase-aligned and 2.63 phase-sensitive.

## G.3 Figure S3

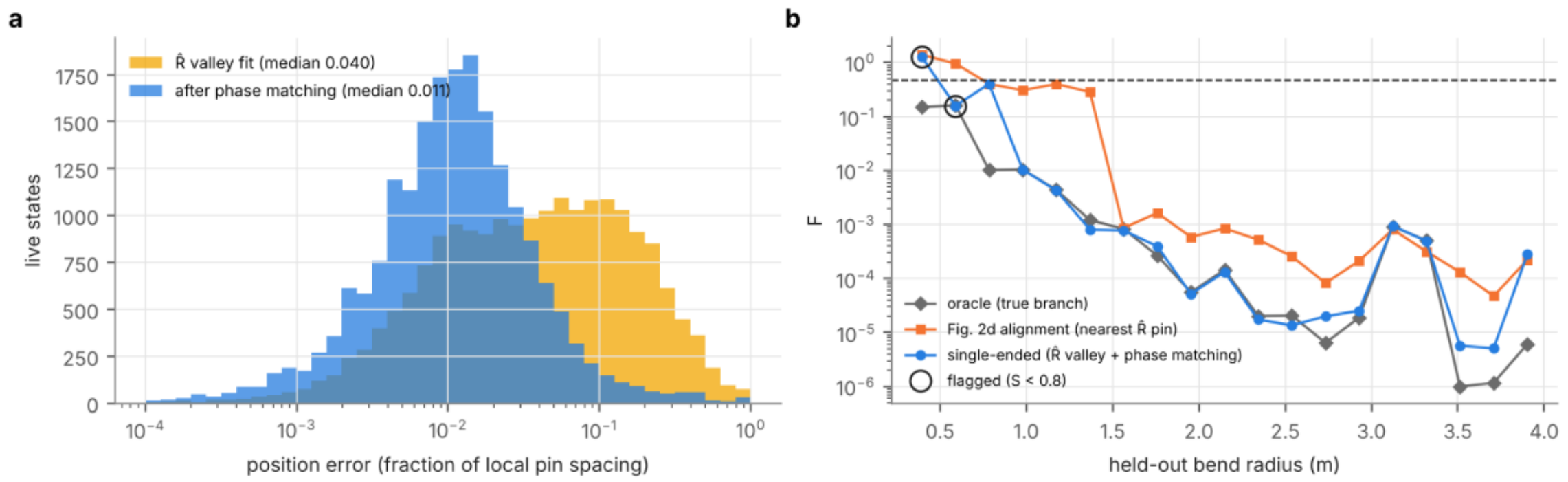

**Fig. S3.** Locating the live state, and the split of Fig. 1d. **a**, Error in the estimated position between pins, as a fraction of the local pin spacing, over all models and pin layouts of Fig. 8a (19,907 live states): R̂ valley vertex (median 0.040) and after phase matching (median 0.011). **b**, Held-out pyMMF bends with every second bend withheld: oracle alignment, the Fig. 2d alignment and the single-ended pipeline; circles, flagged by S < 0.8; dashed line, η = 0.80.

## G.4 Figure S4

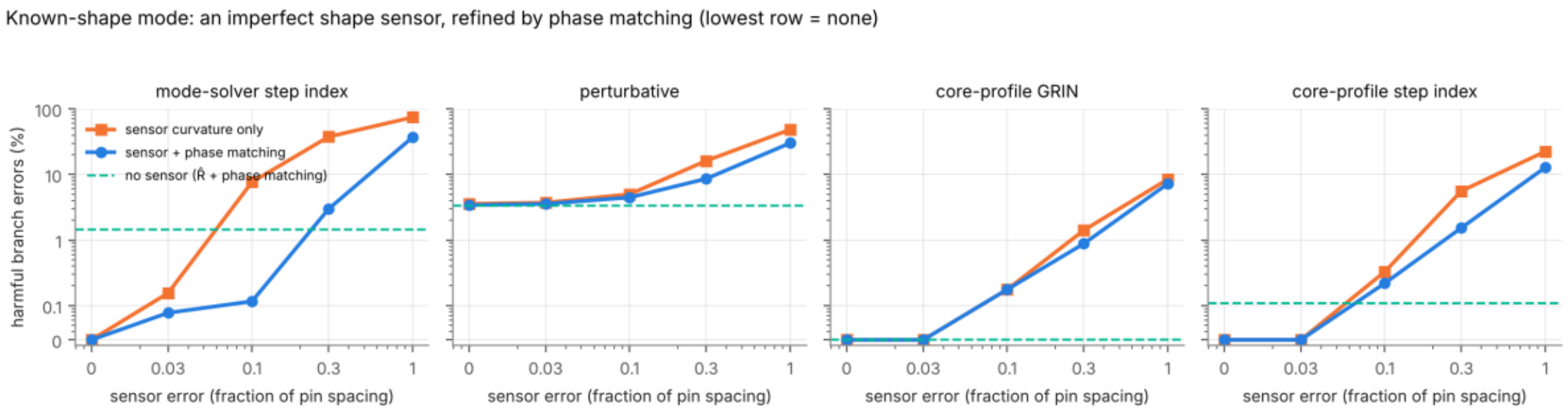

**Fig. S4.** Known-shape mode. Harmful branch errors when a shape sensor supplies the curvature with Gaussian error σ (fraction of the pin spacing), used directly (orange) or as the starting point of phase matching (blue), against the single-ended pipeline with no sensor (dashed), pooled over 16 and 32 evenly spaced and 32 and 48 displaced pins. The lowest row marks zero.

### G.5 Figure S5

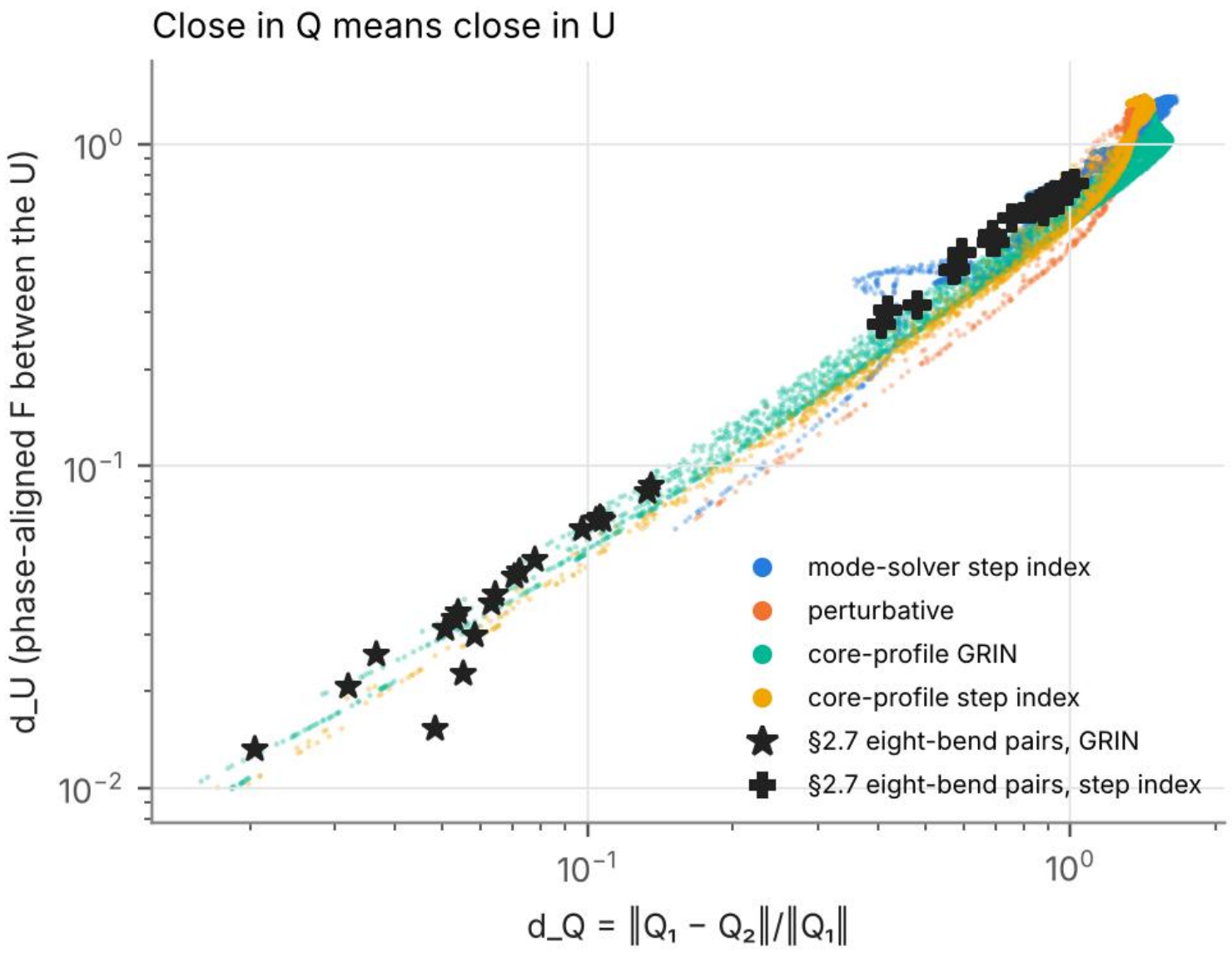

**Fig. S5.** Close in Q means close in U. Phase-aligned distance $d_U$ between transmission matrices against the round-trip distance $d_Q$ for pairs of states in the four single-bend sets (dots, up to 4,000 pairs per set) and for the eight-bend pitch-orthogonal shape pairs of Section 2.7 (stars, GRIN; crosses, step index).

## Appendix H: Robustness and validity checks

Each check below is a self-contained notebook that reruns from the released code and data and saves its own figures.

### H.1 Global phase

A pure global phase leaves every focus unchanged but drives the phase-sensitive $F$ to 2 at 180°; with a real error underneath, the Maréchal threshold is crossed by phase alone at about 25°. All results are therefore quoted with $F$ minimised over a global phase. The effect is a few per cent for the pyMMF benchmark and nil for Takagi Lie-gauge interpolation, which takes its phase from $Q$. It is large where bend-to-bend changes are small compared with the phase they add: thermal-drift errors fall about fourfold, and measured matrices carry phases up to ±174° apart. Phase-aligning the degenerate-block and leave-one-out analyses changes them by under 1%.

## Acknowledgements

GSDG acknowledges support from a UKRI Future Leaders Fellowship (MR/Y034163/1 and MR/T041951/1) and from the Royal Society's Paul Instrument Fund (PIF\R1\261108).

This research was supported by NVIDIA Corporation through the NVIDIA Academic Grant Program and used NVIDIA RTX PRO 6000 Blackwell Max-Q Workstation Edition GPUs.

## Author Contributions

**Terry Wright** conceived the approach: recovering the transmission matrix U of a lossy, non-unitary fibre from a single-ended measurement of the round trip Q and a calibrated reflector matrix R. In this approach, Takagi factorisation and branch alignment of the gauge O map the bend configurations onto a single aligned gauge manifold. Local continuity and tangent-space (Lie-algebra) interpolation then estimate the unknown gauge, from which U is recovered. He proposed reflector de-embedding as a score for identifying candidate transmission matrices from Q, and wrote the initial software demonstrating the approach on pyMMF-simulated fibres. He developed the manuscript from the first draft, extended many of the tests, and checked the results for robustness, including repeating key analyses across simulators and with the global phase removed.

**George S. D. Gordon** developed the approach further and designed the tests presented throughout the paper, including wavelength diversity for determining O and the search for refractive-index profiles that maximise the smoothness of the gauge manifold. He developed the perturbative coupled-mode Hamiltonian models used in Section 2, and wrote the first draft of the manuscript and figures.

**Callum J. Wilson** contributed to acquiring the experimental 5 m multimode-fibre dataset of transmission matrices, intensity reflection matrices and fibre-shape splines across bend configurations.

All authors reviewed and approved the final manuscript.

## Competing Interests

No competing interests

## Data and Code Availability

The code, worksheets and stored results that reproduce every figure and number in this paper, including the simulated pyMMF bend matrices, are archived on Zenodo as takagi-mmf-single-ended-tracking (https://doi.org/10.5281/zenodo.24594249; code under the MIT licence, data and figures under CC BY 4.0). Its CODE_MAP.md file states which code produced each figure and each quoted number. The measured transmission and reflection matrices and fibre shapes are from the public dataset of ref. 29.